\documentclass[twocolumn,twocolappendix]{aastex631}
\usepackage{amsmath}
\usepackage{amssymb}
\usepackage{xcolor}

\accepted{Aug-2022, ApJ}

\shorttitle{Flux Transfer Events}
\shortauthors{Paul, Vaidya \& Strugarek}

\graphicspath{{./}{figures/}}

\begin{document}
\nolinenumbers
\title{A Volumetric Study of Flux Transfer Events at the Dayside Magnetopause}

\author[0000-0003-2287-0368]{Arghyadeep Paul$^*$}
\affiliation{Department of Astronomy Astrophysics and Space Engineering, Indian Institute of Technology Indore,\\Khandwa Road, Simrol, Indore 453552, India}

\author[0000-0001-5424-0059]{Bhargav Vaidya}
\affiliation{Department of Astronomy Astrophysics and Space Engineering, Indian Institute of Technology Indore,\\Khandwa Road, Simrol, Indore 453552, India}
\affiliation{Center of Excellence in Space Sciences India, Indian Institute of Science Education and Research,\\ Mohanpur, Kolkata 741246, India}

\author[0000-0002-9630-6463]{Antoine Strugarek}
\affiliation{Département d’Astrophysique/AIM, CEA/IRFU, CNRS/INSU, Université Paris-Saclay,\\ Université de Paris, F-91191, Gif-sur-Yvette, France}

\correspondingauthor{Arghyadeep Paul}
\email{arghyadeepp@gmail.com}

\begin{abstract}
Localized magnetic reconnection at the dayside magnetopause leads to the production of Flux Transfer Events (FTEs). The magnetic field within the FTEs exhibit complex helical flux-rope topologies. Leveraging the Adaptive Mesh Refinement (AMR) strategy, we perform a 3-dimensional magnetohydrodynamic simulation of the magnetosphere of an Earth-like planet and study the evolution of these FTEs. For the first time, we detect and track the FTE structures in 3D and present a complete volumetric picture of FTE evolution. The temporal evolution of thermodynamic quantities within the FTE volumes confirm that continuous reconnection is indeed the dominant cause of active FTE growth as indicated by the deviation of the P-V curves from an adiabatic profile. An investigation into the magnetic properties of the FTEs show a rapid decrease in the perpendicular currents within the FTE volume exhibiting the tendency of internal currents toward being field aligned. An assessment on the validity of the linear force-free flux rope model for such FTEs show that the structures drift towards a constant-$\alpha$ state but continuous reconnection inhibits the attainment of a purely linear force-free configuration. Additionally, the flux enclosed by the selected FTEs are computed to range between 0.3-1.5 MWb. The FTE with the highest flux content constitutes $\sim$ 1\% of the net dayside open flux. These flux values are further compared against the estimates provided by the linear force-free flux-rope model. For the selected FTEs, the linear force-free model underestimated the flux content by up to 40\% owing to the continuous reconnected flux injection.
\end{abstract}

\section{Introduction} \label{sec:intro}
Magnetic reconnection is a fundamental process occurring in space plasma wherein a rearrangement of topology of the oppositely oriented magnetic field lines leads to an overall magnetic relaxation of the system. Numerous observations, simulations and laboratory experiments have routinely attested to the affiliation of the process with energy conversion and particle acceleration \citep{Yamada_1997, Dal_Pino_2015, Li_2017, Ergun_2020, Dahlin_2020, Arghyadeep_2021}. A seminal paper by  \citet{Dungey_1961} presented the idea that the process of reconnection is a key component of the solar wind-magnetosphere interaction. We now know that reconnection at the Earth's magnetopause facilitates the inflow of mass, momentum and energy into the magnetospheric system. Whether reconnection occurs in a steady continuous fashion, or if the process is inherently intermittent has long been debated. Even though \textit{in-situ} observations of magnetopause reconnection lean  strongly towards the transient nature of the process \citep{Voros_2017, AkhavanTafti_2018,Hoilijoki_2019,Ng_2021}, additional studies have also highlighted that magnetopause reconnection can as well be continuous or quasi-steady for extended periods of time \citep{Frey_2003, Phan_2004, Trattner_2015}. \citet{Strugarek2017} have highlighted the importance of an accurate quantification of the magnetospheric response to the incoming stellar wind in order to assess the dynamical steady state in a more generalised exoplanetary scenario. Studies have further shown that the process of magnetic reconnection itself plays a fundamental role in the formation of a dynamical steady-state magnetosphere\citep{Das_2019}.

Flux transfer events (FTEs) are considered to be the signatures of transient reconnection at the Earth's magnetopause. Multiple studies investigating such structures have concluded that the FTEs form due to patchy localised X-line reconnection at the magnetopause and are then advected away from their generation point \citep{Berchem_1984, Kawano_1996, Daum_2008}. These FTEs manifest in the form of complex helical flux ropes that show up as bipolar signatures in the projection of the magnetic field along a direction perpendicular to the magnetopause boundary layer \citep{SOUTHWOOD1988503,AkhavanTafti_2018}.  It has also been known from in-situ composition data that FTEs can also serve as sites for the mixing of magnetospheric and magnetosheath plasma, further emphasising its importance in the injection of solar wind plasma into the magnetospheric system \citep{Scholer_1982, Paschmann_1982}.

Decades of studies have indeed helped advance the understanding of FTEs and their associated effects on the global magnetospheric system. \citet{Paschmann_1982} concluded that the FTEs had an enhanced interior magnetic field strength, and the total pressure (magnetic + thermal)  inside the FTEs were strongly enhanced compared to the ambient. They therefore inferred that this excess pressure is balanced by the tension of the draped field lines around the FTEs. \citet{Kawano_1997} presented a study on the dependence of the number of FTEs on the interplanetary magnetic field (IMF) configuration concluding that subsolar FTEs mainly form during southward IMF and their east-west motion is controlled by the IMF-$B_y$ . Studies performed by \citet{Kuo_1995} highlighted that physical parameters such as plasma-$\beta$, Mach number, etc., of the solar wind has a weak control over the rate of FTE occurrence. With an increase in observational reports of FTEs, prototypical models were put forward to explain the formation of FTEs such as the \citet{Russell1978} model, the \citet{Lee_1985} model, the \citet{SOUTHWOOD1988503} \& \citet{Scholer_1988} model etc. However, owing to observational limitations and modelling challenges associated with such a multi-scale phenomenon, much has remained elusive. Characteristics such as the azimuthal extent of FTEs,  are not only difficult to estimate, but are nearly impossible to ascertain from in-situ observations due to constraints associated with the relative motion of the spacecraft to the FTE. The trajectory dictates that the spacecraft generally intersects a cross section of an FTE rather than its length which necessitates the utilization of complementary observations for estimating the flux-rope length \citep{Milan_2016, Sun_2019}. 

Numerical simulations have proven to be a functional tool in order to bridge this gap. \citet{Daum_2008} performed a global magnetohydrodynamic (MHD) simulation using the highly acclaimed \textit{Block-Adaptive-Tree-Solarwind-Roe-Upwind-Scheme} (BATS-R-US) code suite and concluded that the modelled results are in good agreement with previous observational and theoretical studies on individual FTE properties. \citet{Dorelli_2009} showed that a dipole tilt is not a necessary condition for FTE formation. Additionally, such MHD simulations of flux transfer events e.g.,  by \citet{Dorelli_2009}, \citet{Cardoso_2013}, have  proven to be useful in deciphering the formation and evolution mechanism of the FTEs. \citet{Sun_2019} have addressed using MHD simulations, the macro-scale characteristics of FTEs in terms of their azimuthal size, location of origin and propagation velocity. Hybrid simulations by \citet{Tan_2011} have further highlighted the formation of a quadrupolar magnetic field structure associated with Hall fields. More recently, the study of FTEs via a two-way coupled magnetohydrodynamics with embedded particle-in-cell code (MHD-EPIC) has made it possible to incorporate kinetic-scale physics into an otherwise global scale magnetohydrodynamic simulation \citep{Chen_2017}. Specifically, simulations by \citet{Chen_2017} have exhibited the multiple X-line origin of FTEs and further highlighted the formation of crescent shaped electron phase-space distributions and the lower hybrid drift instability (LHDI) near the boundary region. Such crescent shaped distributions and LHD waves have also been observed by the MMS probes on multiple occasions \citep{Lapenta_2017, Hwang_2017, Chen_2017, Tang_2019}. 

It is known from observations that the FTEs at the Earth's magnetopause can have a large range of sizes. Cluster surveys, over a period of three years, concluded from a set of 1098 FTEs observed at the Earth's magnetopause that FTE size distributions range within 1000-25000 km with a nominal diameter of 5300 km \citep{Wang_2005,Fermo_2011,AkhavanTafti_2018}. With the advent of the MMS-probes , a larger number of small-scale FTEs closer to the subsolar region could be probed \citep{AkhavanTafti_2018}. The FTEs are believed to grow in size as they are advected away from their generation region on the magnetopause surface which indicates that newly formed FTEs have a smaller size than their `old/mature' counterparts.  The `size' in observations refer to a cross sectional length scale, which is generally the diameter of the FTE. \citet{Akhavan-Tafti_2019b} have determined from the pressure-diameter distribution of a subset of FTEs that the dominant cause of FTE growth is the process of continuous reconnection that feeds in hot exhaust plasma into the flux ropes. As the azimuthal extent of the FTEs cannot be inferred from such observations, the underlying assumption is that the length of the flux rope is constant throughout the evolution process. Their study also addresses multiple features of FTE evolution from a statistical perspective, such as the distribution of plasma moments and forces across the FTE cross sections.  

It has been reported in observations that magnetospheric flux ropes generally appear to be magnetically force-free, i.e the Lorentz force, $\textbf{J}\times\textbf{B}$, is vanishingly small and the magnetic pressure is balanced by the tension force of the twisted field lines \citep{Yang_2022, AkhavanTafti_2018, AkhavanTafti_2019}. The Lepping-Burlaga constant-$\alpha$ flux-rope model ($\textbf{J}=\alpha\textbf{B}$, where $\alpha$ is a constant) has therefore seen widespread use in modelling such transients in order to infer the physical properties of FTEs such as, the size, the core magnetic field strength, the flux content of FTEs, etc. \citep{Burlaga_1988,Slavin_2003,Eastwood_2012,Jasinski_2016,AkhavanTafti_2018, AkhavanTafti_2019}.

Despite being an extensive area of research for over thirty years, many gaps remain in our understanding which could be further bridged by numerical simulations \citep{Sun_2019}. A large obstacle is due to the inherent nature of in-situ observations which are limited to the cross section of the detected FTEs and generally lack information along the flux rope length. The efficacy of studies concerning the FTE volume from the available in-situ data, therefore, becomes dependent on the spacecraft trajectory. Even within the context of numerical simulations, earlier 3-dimensional studies have largely focused on results obtained within a slice through the cross-section of the simulated FTEs in order to directly draw analogies with observations and have omitted any volumetric inferences. This is intensified further due to the fact the FTEs can have an arbitrary shape occupying an irregular volume. Our study aims to fill this gap, and, for the first time to our knowledge, we present studies of FTEs powered by a complete volumetric detection of these arbitrary structures in 3 dimensions. This volumetric detection technique has also facilitated the tracking of the FTE volumes over time to investigate the temporal evolution of various physical properties of the FTEs. We leverage this method to investigate the evolution of thermodynamic properties within the FTE volume in order to assimilate the mechanisms of FTE growth. We further probe into the temporal evolution of the magnetic properties of the FTEs. In particular, we investigate the time evolution of perpendicular currents within the FTE volume and further contrast our FTEs to the analytical constant-$\alpha$ flux-rope model. Finally, we compare the flux content of the simulated FTEs with estimates obtained from this analytical model and further quantify the total dayside closed and open magnetic flux within our simulation.

The article is organised as follows. Section \ref{sec:Setup} describes the model setup, the numerical framework along with the initial conditions and the volumetric FTE detection technique used in this study. Section \ref{sec:Results} describes in detail, the principal findings of this study and section \ref{sec:Discussion_and_summary} summarizes the article and presents additional discussions and future prospects relevant to the study. 

\section{Methodology}\label{sec:Setup}
In order to model the interaction of the ambient solar wind with an Earth-like planetary magnetosphere, we perform numerical simulations with adaptive mesh refinement (AMR) on a Cartesian mesh.

\subsection{Numerical Approach}
The numerical approach involves solving a set of resistive MHD equations in 3 dimensions using the MHD module of the PLUTO code \citep{Mignone_2012}. The module solves a set of conservation laws in the form of single-fluid MHD equations given as:
\begin{equation}\label{MHD_eqs}
\begin{aligned}
    \frac{\partial \rho}{\partial t} + \nabla .(\rho \mathbf{v})  &=  0 \\
    \frac{\partial (\rho  \mathbf{v})}{\partial t} + \nabla . \left[\rho \mathbf{v}\mathbf{v} - \mathbf{B}\mathbf{B}\right] + \nabla \left( p + \frac{\mathbf{B}^2}{2} \right)  &= 0 \\
    \frac{\partial \mathbf{B}}{\partial t} + \nabla \times (c\mathbf{E}) &= 0 \\
    \frac{\partial E_{t}}{\partial t} + \nabla . \left[ \left( \frac{\rho\mathbf{v}^2}{2} +\rho e + p\right)\mathbf{v}  + c\mathbf{E}\times \mathbf{B} \right] &= 0
\end{aligned}
\end{equation}
where $\rho$ is the mass density, $\mathbf{v}$ is the gas velocity, $p$ is the thermal pressure and $\mathbf{B}$ is the magnetic field. A factor of 1/$\sqrt{4\pi}$ has been absorbed in the definition of $\mathbf{B}$. $E_{t}$ is the total energy density which can be described as:
\begin{equation}\label{eq:energy}
    E_{t} = \rho e + \frac{\rho \mathbf{v}^{2}}{2} + \frac{\mathbf{B}^2}{2}
\end{equation}

An \textit{ideal} equation of state provides the closure as $\rho e = p/ (\gamma -1)$ wherein, $\gamma$ is the ratio of specific heats having a value of 5/3. The electric field, \textbf{E}, is composed of a convective and a resistive component and is defined as:
\begin{equation}\label{eq:induction_eq}
    c\mathbf{E} = -\mathbf{v}\times \mathbf{B} + \frac{\eta}{c} \mathbf{J}
\end{equation}
where $\eta$ is the resistivity and $\mathbf{J}= c\nabla\times\mathbf{B}$ is the current density. To detach the dependence of the reconnection process on the numerical resistivity of the system, we incorporate an explicit resistivity in the domain having a value of $\rm \eta /\mu_{0}= 1.27 \times 10^{9} (m^2/s)$. It is expected that the inclusion of an explicit resistivity will broaden the current layer at the magnetopause \citep{Komar_thesis}. The chosen explicit resistivity was motivated by its outcome that the full width at half maxima (FWHM) of the current layer be resolved by no less than 6 to 7 grid cells. This, in turn, also ensures that the entire span of the magnetopause current sheet is always resolved by at least 12 to 14 grid cells. Due to the high computational costs, we have refrained from performing parameter dependence studies on the value of this explicit resistivity. The flux computations have been performed with the second order accurate Harten-Lax-vanLeer (HLL) solver and the solenoidal constraint ($\nabla\cdot\textbf{B}$ = 0) is enforced by coupling the induction equation to a  Generalised Lagrange Multiplier (GLM) and solving a modified set of conservation laws in a cell-centered approach \citep{Dedner_2002, Mignone_2012}. 

Additionally, we have adopted a formalism where the total magnetic field $\mathbf{B}$ inside the domain is treated in a split configuration given as 
\begin{equation}\label{eqn:split_field}
    \mathbf{B}(x,y,z,t) = \mathbf{B}_{0}(x,y,z) + \mathbf{B}_{1}(x,y,z,t)  
\end{equation}
where $\textbf{B}_0$ is a curl free, time invariant, background magnetic field and $\mathbf{B}_1$ behaves as a deviation. For such a configuration, the energy depends only on the deviation $\mathbf{B}_{1}$ which turns out to be computationally convenient when dealing with magnetically dominated systems.
\subsection{Initial and Boundary Conditions}
The above set of resistive -MHD equations are solved in a Cartesian box enclosed by  $\rm -30 R_{E}\leq X \leq 130 R_{E}$, $\rm -100 R_{E}\leq Y \leq 100 R_{E}$ and $\rm -100 R_{E}\leq Z \leq 100 R_{E}$, where $\rm R_E$ denotes the Earth radius. The domain has a base resolution of 64 $\times$ 80 $\times$ 80 grid cells which is refined using a block structured AMR strategy. The AMR strategy only refines the regions where a higher resolution is required (e.g, the dayside magnetopause) keeping the remaining computational grid coarse. This strategy therefore proves to be extremely beneficial in cases with a high disparity of scale sizes and saves a significant amount of computational expense. The details of the AMR implementation in PLUTO can be found in \citet{Mignone_2012}. The mesh is adaptively refined up to four refinement levels in a hierarchical manner where the first two and the last two AMR levels have a refinement ratio of 2x and 4x respectively. This leads to an effective resolution of 4096 $\times$ 5120 $\times$ 5120 which sets the smallest grid size at $\Delta x = \Delta y = \Delta z =$ 0.039$\rm R_E$. The refinement criterion is set to be the current density $\mathbf{J}$ which dictates that regions with sharp gradients in $\mathbf{J}$ will be recursively refined to finer resolutions.

An internal boundary (IB) is set at radii $\rm r \leq 4R_{E}$, inside which the values of the MHD variables are kept fixed. A thermal pressure of 6.69 $\times 10^{-3}$nPa and a density proportional to the local magnetic field strength is prescribed inside the IB. The IB acts as a sink and any MHD variable that flows in to the IB is overwritten by these fixed values. The magnetic field is considered to be purely dipole like inside the IB and the aforementioned density profile prevents the Alfv\'en speed from attaining large values at the IB which assists in limiting the simulation time-step. Plasma-$\beta$ is defined as the ratio of the gas pressure to the magnetic pressure. In low $\beta$ plasmas (generally found in regions of high magnetic field strength in global MHD simulations), negative pressure values may arise due to truncation errors while calculating the thermal pressure, generally expressed as a difference of two very large numbers (following equation \eqref{eq:energy}). To ensure that such unphysical values of pressure do not arise in the domain, the equation of entropy is also added to the set of conservation laws given by equation \eqref{MHD_eqs}. Even though the energy equation is used everywhere in the domain, the pressure is computed from the entropy after AMR operations such as coarse-to-fine prolongations and restrictions \citep{Mignone_2012}. This procedure is one of the many techniques that can be used to ensure pressure positivity in simulations where the magnetic energy is the dominant contribution to the total energy density \citep{Lyon_2004, Ridley_2010, Li_2021}. 

To emulate the dipolar magnetic field of an Earth-like planet, the configuration for $\mathbf{B_{0}}$ in equation \eqref{eqn:split_field} is prescribed as a magnetic dipole placed at the origin of this box having an equatorial field strength of $3 \times 10^{-5}$ T at 1$\rm R_E$. The axis of the dipole is tilted by 5$^{\circ}$ with respect to the Z-axis with its south-pole leaning towards the Sun. This small tilt helps provide a preferred direction to the propagation of transient reconnection events such as FTEs on the magnetopause surface. The entire domain is initially filled with a very low density of 0.5 amu cm$^{-3}$. Thereafter, a solar wind inflow is prescribed at the left X-boundary that has a modest plasma inflow speed of $v_{sw}=$ 400kms$^{-1}$, a density of $\rho_{sw}$= 5 amu cm$^{-3}$ and a pressure of $P_{sw}$= 3 $\times 10^{-2}$nPa. An interplanetary magnetic field with components $B_{y(sw)}= 5$nT and $B_{z(sw)}= -5$nT is also prescribed at the inflow boundary which then propagate into the domain with the solar wind. The input conditions correspond to typical solar wind parameters at 1AU \citep{GOSLING2014,Klein_2019} and similar inflow conditions have previously been used in the literature for numerical studies on FTEs \citep{Sun_2019}. Once the inflow in established, the sonic ($\mathit{M}_{S}$) and the Anfv\'enic Mach numbers ($\mathit{M}_{A}$) are calculated to be $\mathit{M}_{S} \sim$ 4.9 and  $\mathit{M}_{A} \sim$ 5.9 respectively. The remaining five boundaries are set to have Neumann boundary conditions which dictates that all MHD variables have zero gradient across these boundaries.

\section{Results}\label{sec:Results}
\subsection{Initial Evolution Characteristics}\label{sec:init_evol_char}
\begin{figure*}
    \centering
    \includegraphics[width=\textwidth]{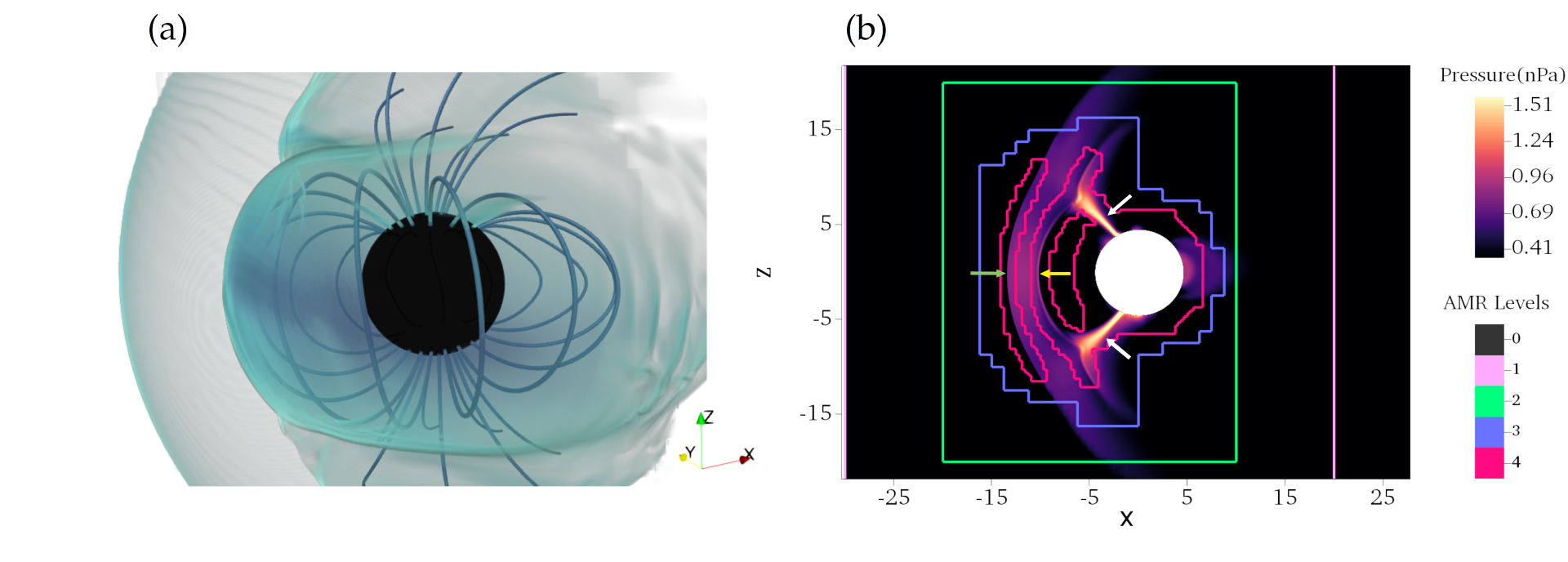}
    \caption{Panel (a) shows a volumetric rendering of the current density magnitude ($\lvert$\textbf{J}$\rvert$) profile highlighting the simulated magnetosphere. The black sphere represents the internal boundary at $\rm r = 4R_{E}$. Panel (b) shows the AMR levels after the day-side magnetopause attains a nearly steady state. Level `0' corresponds to the entire base grid and the coloured boxes represent the remaining four refinement levels as per the legend on the bottom right. Background pseudo-color in panel (b) indicates thermal pressure in nPa. The green arrow in panel (b) denotes the bow-shock, the yellow arrow highlights the magnetopause and the white arrows denote the polar-cusps.}
    \label{fig:vol_plus_AMR_levels}
\end{figure*}

As the simulation starts, the solar wind plasma flows in from the left X-boundary and interacts with the magnetic field of the dipole compressing it on the day side and stretching it out on the night side forming a paradigmatic magnetosphere as seen in panel (a) of figure \ref{fig:vol_plus_AMR_levels}. The panel shows a volumetric rendering of the simulated magnetosphere 
\hspace*{1.0cm}

During the early phase of the simulation, the evolving mesh refines the regions of strong gradients in current density as directed by the AMR refinement criteria. The day side magnetopause region is expected to attain a local maxima of the current density due to the southward IMF of the incoming flow. We find that this region is indeed adaptively refined to the finest level and in due course, the polar cusps are refined as well. The magnetopause eventually attains a nearly steady state after $t\sim960$s (in units of physical time) as shown in panel (b) of Figure \ref{fig:vol_plus_AMR_levels}. In Figure \ref{fig:vol_plus_AMR_levels}, the resolution of the base grid ($\Delta x= \Delta y= \Delta z=  2.5 R_{E}$) in each of the three dimensions is refined to $\rm 1.25 R_{E}$, $\rm 0.625 R_{E}$, $\rm 0.156 R_{E}$ and $\rm 0.039 R_{E}$ by the AMR levels 1, 2, 3 and 4 as denoted by the legend. The white circle centered at the origin represents the internal boundary. 

First, we identify the day-side magnetopause in our simulated magnetosphere as an isosurface of the maximum value of the current density \textbf{J} within $\rm -8 R_{E}\leq Y \leq 8 R_{E}$ and $\rm -5.5 R_{E}\leq Z \leq 5.5 R_{E}$. This exercise has been done during a quiet time when the surface is seemingly stationary and devoid of any transient phenomena. As a validation step, we then compared this detected surface to the analytical model of \citet{Cooling_2001} which considers the magnetopause to be a paraboloid of revolution. We confirm that the two are well in agreement. In particular, we compared the X-coordinate of a sample of more than $10^{5}$ points from the simulated magnetopause surface to their corresponding analytical positions and found an excellent agreement with a correlation coefficient of 0.99 and a root mean square (RMS) deviation of $\rm \sim$0.1$\rm R_E$ across the entire surface. The detected magnetopause, in addition to serving as a validation step, also has multiple additional usage as will be evident in the following sections.

\subsection{FTE Generation and Signatures}
Due to the presence of a significant magnetic shear as a consequence of the southward IMF, the dayside magnetopause exhibits a local maxima of the current density (\textbf{J}) and thereby turns out to be a prime location for magnetic reconnection. The following paragraphs describe the process of magnetic reconnection from a magnetohydrodynamic perspective in order to shed light into how the process occurs within a resistive MHD scenario. Magnetic reconnection occurs in resistive-MHD simulations due to the presence of an explicit resistivity in the induction equation (equation \ref{eq:induction_eq}) which facilitates dissipation. Within the context of MHD, regions of high current density locally enhances the contribution of the second term in equation \ref{eq:induction_eq} leading to the formation of a diffusion region at ion scales where the frozen-in constraint of the magnetic field lines breaks down.
\begin{figure*}
    \centering
    \includegraphics[width=\textwidth]{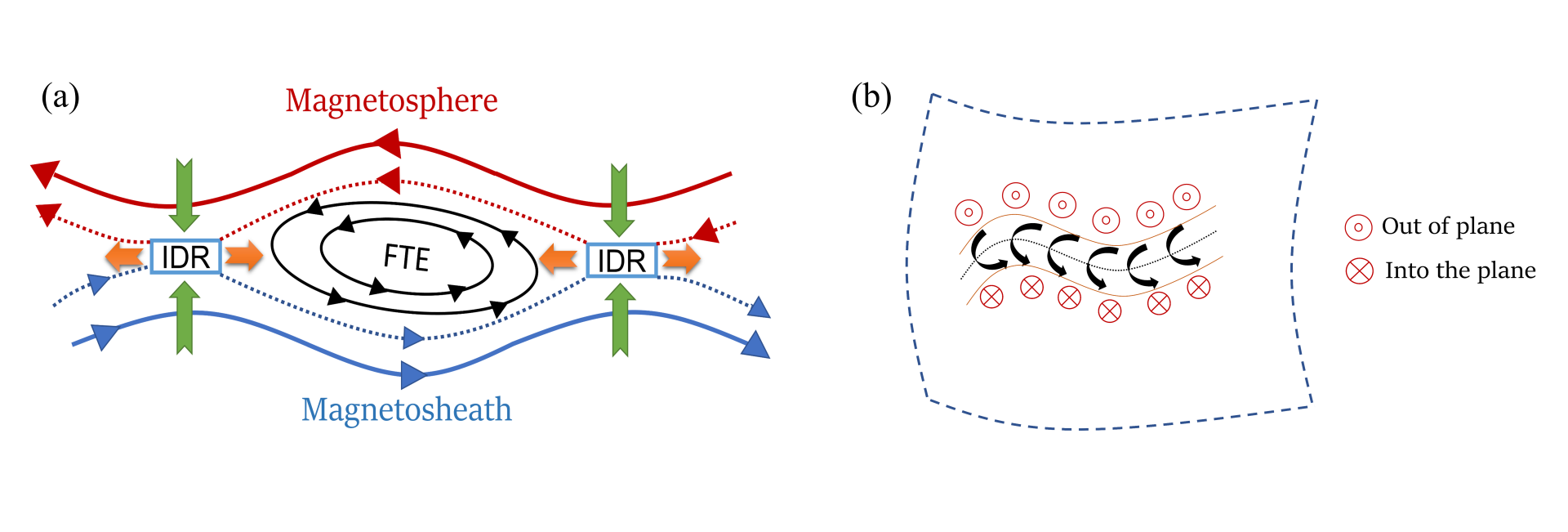}
    \caption{Panel (a) shows a 2D cross sectional slice (slice across the FTE axis) of a typical FTE forming at the magnetosphere-magnetosheath boundary layer. The two ion diffusion regions are marked as `IDR'. The reconnection inflows and outflows are denoted by the green and orange arrows towards and away from the IDR respectively. Panel (b) shows a typical slice of an FTE along the FTE axis on the boundary layer. The black dotted line corresponds to the FTE axis and the black solid arrows denote the magnetic field directions of a helical field line intersecting the surface. The red dotted circles show that the regions above the FTE axis have magnetic field lines moving out of the boundary layer whereas the red crossed circles denote that the regions below the FTE axis have magnetic field lines moving into the boundary layer.}
    \label{fig:schematic_reconnection}
\end{figure*}

Panel (a) of figure \ref{fig:schematic_reconnection} shows a cross sectional slice (2D representation) across the axis of a typical FTE formed due to magnetic reconnection at the magnetopause. The magnetospheric (dipolar field lines) are marked in solid red color whereas the solar wind magnetic field lines on the magnetosheath side are marked in solid blue color. Two ion diffusion regions marked as `IDR' are shown in the schematic. These IDR are the sites where the magnetic field lines of the magnetopause and the magnetosheath reconnect (dotted lines of the corresponding colors). As represented in a 2D projection, the change in topology of the magnetic field lines within the IDR causes a loop like magnetic field configuration in the region between the two IDR marked in solid black field lines. The green arrows pointing into the IDR and the orange arrows pointing away from the IDR represents the reconnection inflows and exhausts respectively. A portion of the hot plasma exhausts from the IDR creates a plasma bulge at the boundary layer. Such a plasma bulge in addition to the characteristic loop like magnetic field topology constitutes an FTE. In 3D however, the magnetic field topology is generally more complex. The IDR can extend in the third dimension (into and out of the plane in panel (a) of figure \ref{fig:schematic_reconnection}) up to an arbitrary length and the loop like magnetic field lines then exhibit the configuration of a helical rope called a flux-rope. Such sporadic patchy reconnection at the magnetopause boundary layer gives rise to multiple FTEs which materialize in the form of transient flux-ropes.

Panel (b) shows a slice of a typical flux rope along its axis on the boundary layer (blue curved surface) where the black dotted line represents the flux rope axis and the solid black arrows represent the field line configuration of the portion of the flux rope above the boundary layer. For such a slice, any constituent helical magnetic field line of the flux rope intersects the boundary layer twice. The magnetic field lines in the upper half of the flux rope axis is seen to be moving out of the boundary layer whereas the field lines on the lower half is seen to be moving into the boundary layer. \textit{In-situ} probes generally measure the $\rm B_N$ component at the boundary layer which is the component of the total magnetic field perpendicular to the boundary layer. Typical \textit{in-situ} FTE detections therefore constitute of a bipolar signature in the $\rm B_N$ component as the FTE passes through the probe (negative due to the lower portion of the FTE and positive due to the upper portion or vice-versa). Such a  bipolar signature can also be correspondingly identified in the $\rm B_X$ component of the magnetic field in Geocentric Solar Magnetospheric (GSM) coordinate system. Though one must note that modelling smaller scale reconnection signatures such as the quadrupolar Hall-fields or lower hybrid drift waves in addition to particle distribution functions in reconnection regions require more sophisticated and computationally intensive numerical frameworks such as Hall-MHD, hybrid-Vlasov or particle-in-cell, modelling approaches \citep{Chen_2017, Hoilijoki_2019, Tafti_2020_vlasiator}.

\begin{figure*}
    \centering
    \includegraphics[width= \linewidth]{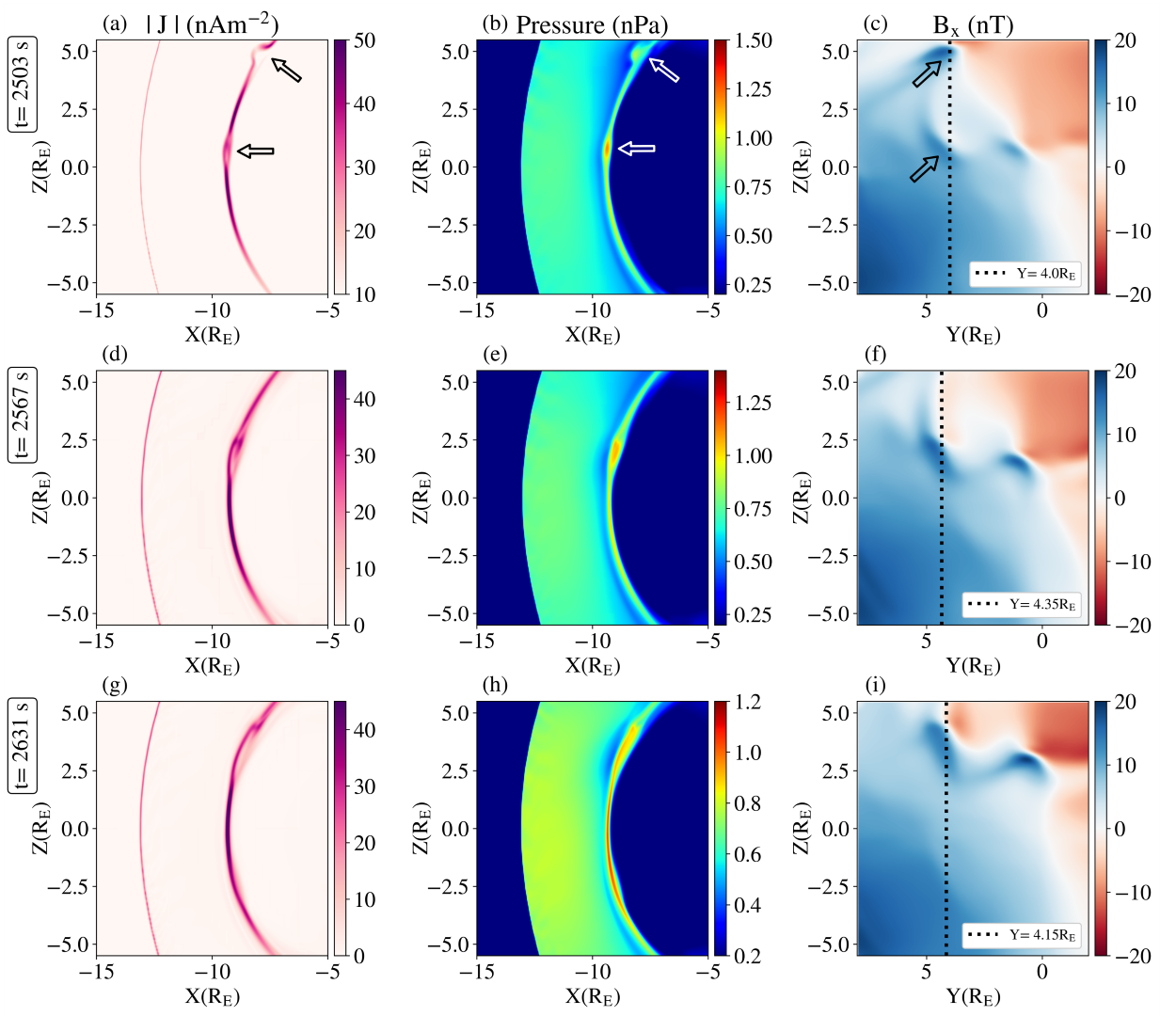}
    \caption{Plots of the magnitude of current density (panels (a), (d) and (g)), thermal pressure (panels (b), (e) and (h)) and a projection of the $\rm B_X$ component at the magnetopause surface (panels (c), (f) and (i)) highlighting the formation and advection of FTEs in our simulation. The rows denote different times, wherein, panels (a), (b) and (c) are at t= 2503 s, panels (d), (e) and (f) are at t= 2567 s and panels (g), (h) and (i) are at t= 2631 s. The arrows in panels (a), (b) and (c) represent the FTE positions for clarity. The dotted line on the plots of the third column represent the Y coordinate where the first and second columns have been sliced.}
    \label{fig:evolution_of_FTE_J_PRS_BX}
\end{figure*}

\begin{figure}
    \centering
    \includegraphics[width= 1.0\columnwidth]{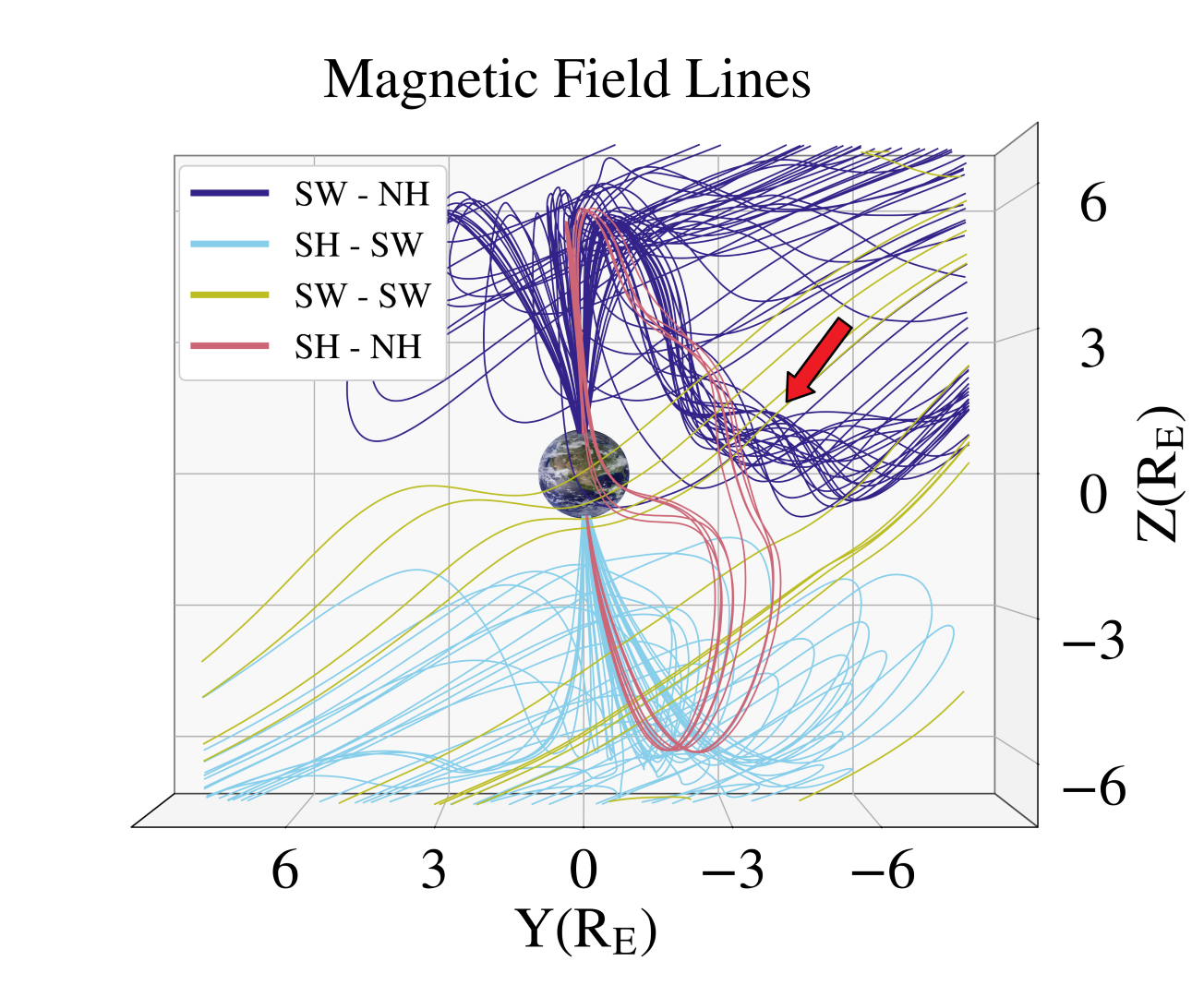}
    \caption{A plot of the magnetic field streamlines in the domain showing different connectivities at t$ \rm \sim3364s$. The field lines are color coded as per the legend where `SW', `NH' and `SH' represent `Solar Wind', `Northern Hemisphere' and `Southern Hemisphere' respectively. A helical flux rope (FTE-3, see section \ref{sec:FTE_det_validation} for nomenclature) can be seen in the region marked by the arrow.}
    \label{fig:streamline}
\end{figure}

Figure \ref{fig:evolution_of_FTE_J_PRS_BX} is a depiction of the temporal evolution of FTEs that form in our simulation. The subplots show a zoomed-in view to distinctly feature the regions of interest. The columns represent different quantities whereas the rows represent different time snapshots. The corresponding physical times are encased in boxes to the left of each row in Figure \ref{fig:evolution_of_FTE_J_PRS_BX}. Panels (a) and (b)show the X-Z slices at y=4.0. Panel (a) shows the magnitude of current density $\rm |\textbf{J}|$ highlighting the bow shock and the magnetopause surface. One can also see two distinct FTEs on the magnetopause surface in panel (a) denoted by the arrows. The FTEs can also be similarly seen in panel (b) which represents the pressure on the same slice as panel (a). Panel (c) shows a projection of the
$B_x$ component on the magnetopause surface. The $B_x$ component shows bipolar features due to the specific magnetic field configuration within the FTEs. One can see these signatures corresponding to the two FTEs of panels (a) and (b) at the location marked by the arrows in panel (c). The dotted vertical line in panel (c) marks the Y-coordinate corresponding to the slices shown in panels (a) and (b). The next row (panels (d), (e)and (f)) are plotted after a time gap of $\rm \sim 64$ seconds. The FTE that was located at the top in panels (a), (b) and (c) have now moved out of the plot and the other FTE is seen to have grown in size. It is also seen that the bipolar $B_x$ signature of the other FTE has strengthened as well. The bottom row is after another 64 seconds which shows that this FTE has now significantly grown in size and one can also see the initial stages of a new FTE forming at $\rm z\sim -2.5$. Such transient features are persistent throughout our simulation and remains in the region resolved by the finest grid size for a significant amount of time which allows for a detailed study of their evolution. We note here for clarity that the FTEs shown in figure \ref{fig:evolution_of_FTE_J_PRS_BX} are a representative of the multiple transient features produced in our simulations and are separate from the ones that have been extensively analysed later in this paper. We also highlight the fact that the prescription of a southward IMF at the inflow boundary and the use of an explicit resistivity to imitate the dissipation of magnetic energy in regions of high magnetic shear are sufficient conditions to initiate magnetic reconnection leading to the production of FTEs within our simulation.
Figure \ref{fig:streamline} is a representation of the magnetic field line connectivities near the magnetopause surface at t$\rm \sim3364s$. To avoid overcrowding the plot, we have only plotted a small fraction of the traced magnetic field lines. The field line connectivities show clear signatures of magnetic reconnection wherein the incoming solar wind magnetic field lines and the Earth's magnetospheric field lines reconnect to produce magnetic fields that are connected to the solar wind at one end and the ionosphere/ northern or southern hemisphere at the other end. These lines are labelled as `SW - NH' and `SH - SW' denoting `Solar Wind to Northern Hemisphere' and `Southern Hemisphere to Solar Wind' connectivities respectively. One can also see a glimpse of a helical magnetic flux rope (dark blue field lines) towards the right side of the picture as marked by the arrow. This flux rope is in fact a depiction of the early stages of one of the FTEs which has been analysed later in this paper (namely FTE-3, see the following section for nomenclature).

\subsection{FTE detection: Validation}\label{sec:FTE_det_validation}
Due to the lack of control over the exact location on the magnetopause where the reconnection occurs in the simulation, the FTEs generally tend to have an arbitrary extent along each coordinate which makes it challenging to determine their complete 3-dimensional structure. In this study, we have used an agglomerative hierarchical structure finding algorithm to perform volumetric detection of the FTEs within our simulation. A detailed methodology of the volumetric detection has been described in appendix \ref{ss:detection methodology} for completeness. Using this process described in appendix \ref{ss:detection methodology}, we have detected and tracked several FTEs moving along the magnetopause surface with time. From the tens of detected FTEs, we select three FTEs, named FTE-1, FTE-2 and FTE-3, for a further detailed analysis. The FTEs were chosen so that they originate and travel along different directions on the magnetopause surface. This is to ensure a broader coverage of the magnetopause and to ascertain that any bias on the FTE properties arising due to their location could be eliminated. One of the principal magnetic field signatures for FTE identification used in in-situ observations is the bipolar $B_N$ component of the magnetic field \citep{Russell1978, Jasinski_2016}. The $B_N$ component, being a part of the boundary normal L-M-N coordinate system, represents the projection of the total magnetic field along a direction that is normal to the magnetopause surface. All the FTEs formed in our simulations universally show such distinct bipolar $B_N$ signatures.
\begin{figure*}
    \centering
    \includegraphics[width= 1.0\textwidth]{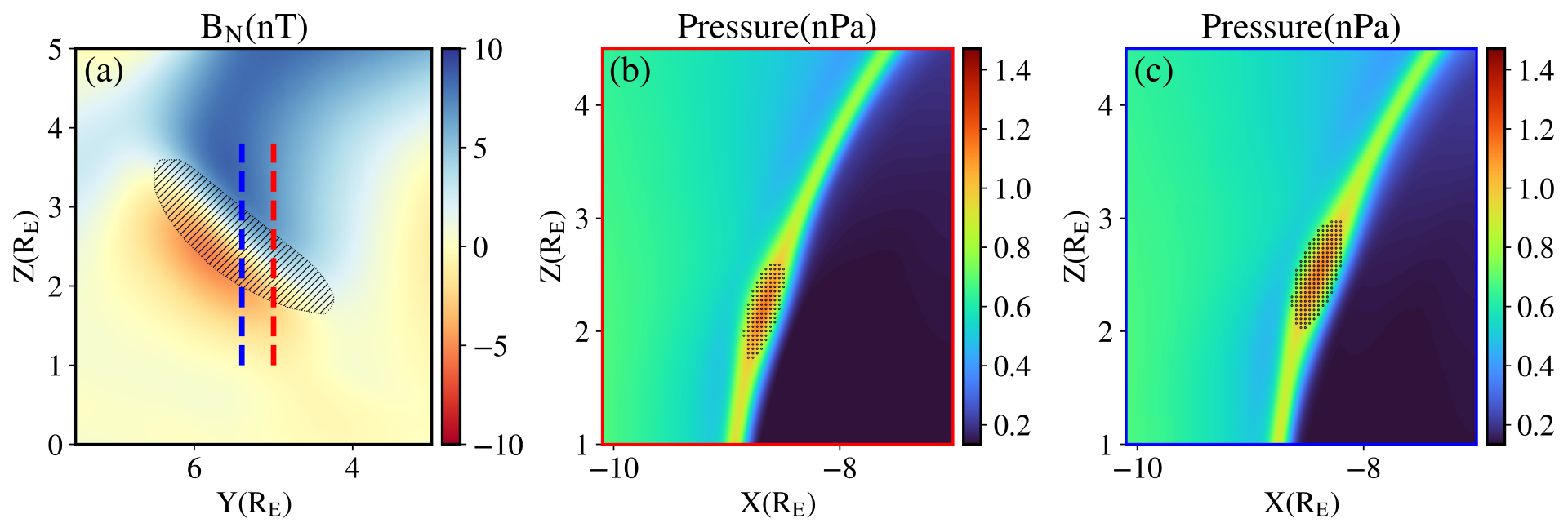}
    \caption{A plot showing slices of FTE-2 detected using the structure finding algorithm. Panel (a) shows the $B_N$ component (sign reversed) of the magnetic field on the magnetopause surface where the shaded region represents the Y-Z projection of the detected FTE volume. The red and blue dashed lines in panel (a) denote the locations along which the Y-slice is taken for the panels (b) and (c) respectively. The panels (b) and (c) show the thermal pressure in the X-Z plane slicing the FTE at y= 5.0 $\rm R_E$ and y= 5.4 $\rm R_E$ respectively. The shaded region in both these panels represents the slice of the FTE along the corresponding y-values.}
    \label{fig:detection}
\end{figure*}
Panel (a) of Figure \ref{fig:detection} shows in hatched line shading, a projection of the azimuthal extent of the detected volume of FTE-2 obtained from the structure detection algorithm, superimposed on the $B_N$ component at the magnetopause surface. The detected structure is seen to span the entire azimuthal extent (the extent on the Y-Z plane) of the bipolar structure (red-blue junction) correctly. Two slices across this FTE along the red and blue dashed lines in panel (a) are shown in panels (b) and (c), where, the shaded regions correspond to the detected FTE structure. Such a validation exercise has been carried out for each of the three FTEs for each time-step by visually comparing the detected structure to the $B_N$ component signature in the X-Z plane as well as along multiple X-Z slices to ascertain that the entirety of the detected structures indeed lie within the FTE thermal pressure bulge. For all the FTEs detected, our volumetric identification is found to provide a fairly accurate detection within a margin of at most two grid cells inside the FTE structure. We note here that on a closer look at panel (a) of figure \ref{fig:detection}, one might posit that there exists positive and negative $B_N$ regions that lie outside the detected FTE volume, however, a detailed inspection of panels (b) and (c) reveals that such regions at the top and bottom of the FTE slice and are in-fact the transition between the ellipsoidal FTE bulge and the magnetopause layer and therefore, excluding these segments, in reality, only isolates the cardinal regions to be the dominant portion of the detected FTE structure. We also note that towards the later stages of FTE evolution, the thermal pressure within the FTE drops as one moves towards the flux rope core. Such a pressure profile has also been seen in observations\citep{AkhavanTafti_2019}. This variation is, however, less prominent during the early stages of FTE evolution.

\begin{figure}
    \centering
    \includegraphics[width= 1.0\columnwidth]{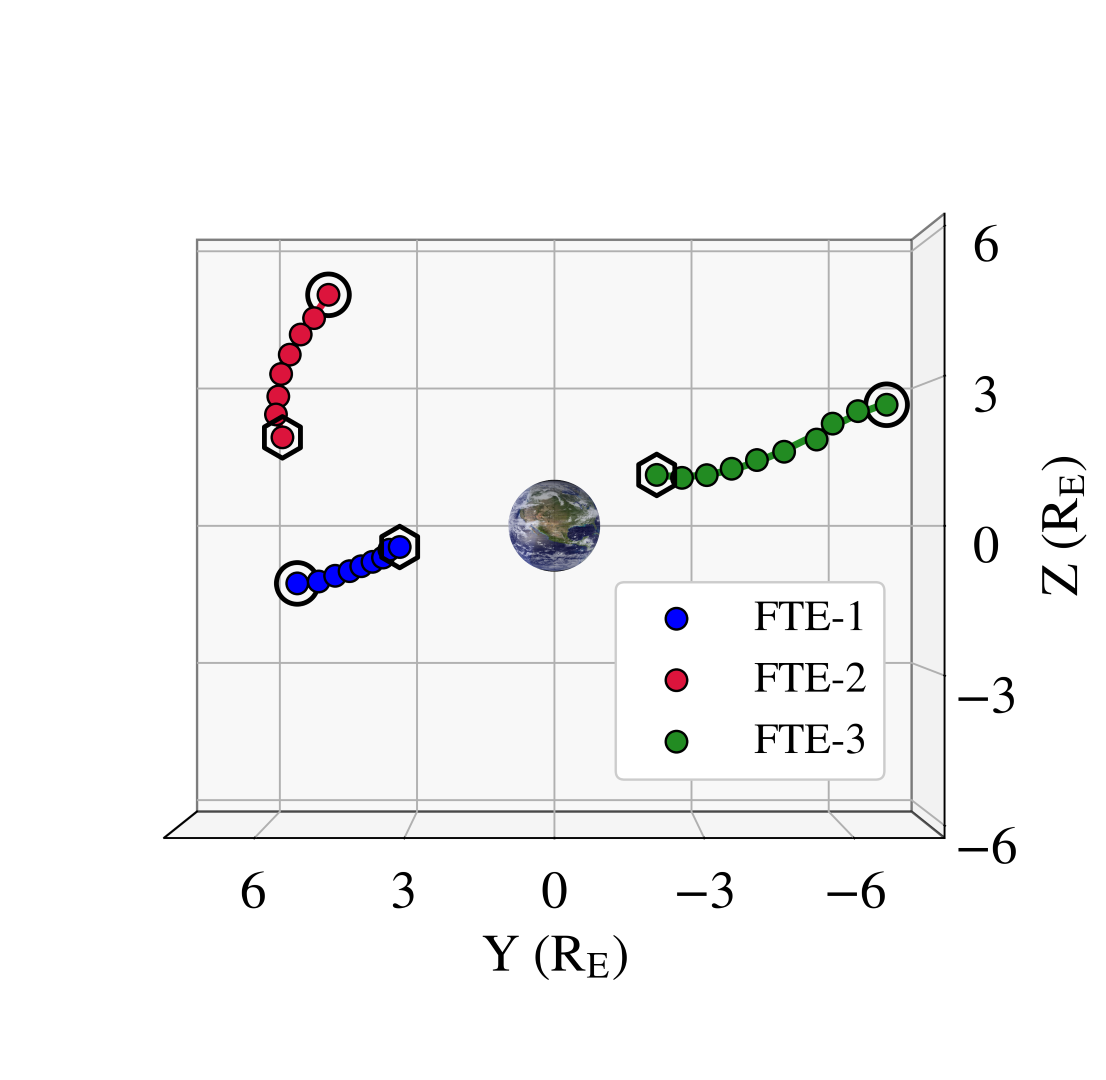}
    \caption{A scatter plot of the trajectory of the centers of mass of the three selected FTEs showing their positions and motion. The points enclosed within the hexagons represent the initial position of the corresponding FTEs whereas the encircled points denote the final location. The sphere at the center is a representative Earth.}
    \label{fig:center_of_mass}
\end{figure}

Figure \ref{fig:center_of_mass} shows the motion of the centers of mass of the selected FTEs with time, as seen when looking from the Sun towards the magnetopause. The FTEs travel along the magnetopause surface which dictates that their motion along the X-direction is governed by the curvature of the magnetopause surface which is insignificant within the bounds of the plot. The dominant motion of the FTEs is therefore in the Y-Z plane. The center of mass motion of the three FTEs can be considered as a representative of the general motion of the FTE volumes along the magnetopause. The multiple scatter points for each FTE corresponds to the different time snaps over which the FTEs were detected and tracked.

Further details of the selected FTEs have been summarised in Table \ref{Table:FTE} where t$_{start}$ is the global time (in seconds) when the FTEs were first detected by the algorithm. The column named $\Delta$t represent the time duration over which each FTE was tracked, the average speed of the FTEs ($\langle$v$\rangle$) in km s$^{-1}$ is shown in the fourth column and the fifth column gives the general location of the selected FTEs. The average speed has been estimated from the total distance travelled by the FTEs in the time interval $\rm \Delta t$.  
\begin{center}
\begin{table}
 \centering
 \begin{tabular}{|c|c|c|c|c|}
 \hline\hline
 Name  & t$\rm_{start}$ (s) & $\Delta$t (s) & $\langle$v$\rangle$ (km s$^{-1}$) & Location  \\ [0.8ex]
  \hline
  FTE-1 &  1371   & 143.5  &    126.9  &  South-West \\ [0.8ex]
  FTE-2 &  2806   & 127.5  &    191.3  &  North-West \\ [0.8ex]
  FTE-3 &  3332   & 159.4  &    247.7  &  North-East \\ [0.8ex]
\hline\hline
 \end{tabular}
\caption{The table summarizes the names, times when the FTEs were first detected in the simulation (t$\rm _{start}$), the time duration ($\rm \Delta t$) over which the FTEs were tracked, the average speed ($\langle$v$\rangle$) and the general location of the selected FTEs.}
\label{Table:FTE}
\end{table}
\end{center}
\subsection{Thermodynamic Properties}\label{sec:Thermodynamic_props} 
\begin{figure}
    \centering
    \includegraphics[width= 0.9\columnwidth]{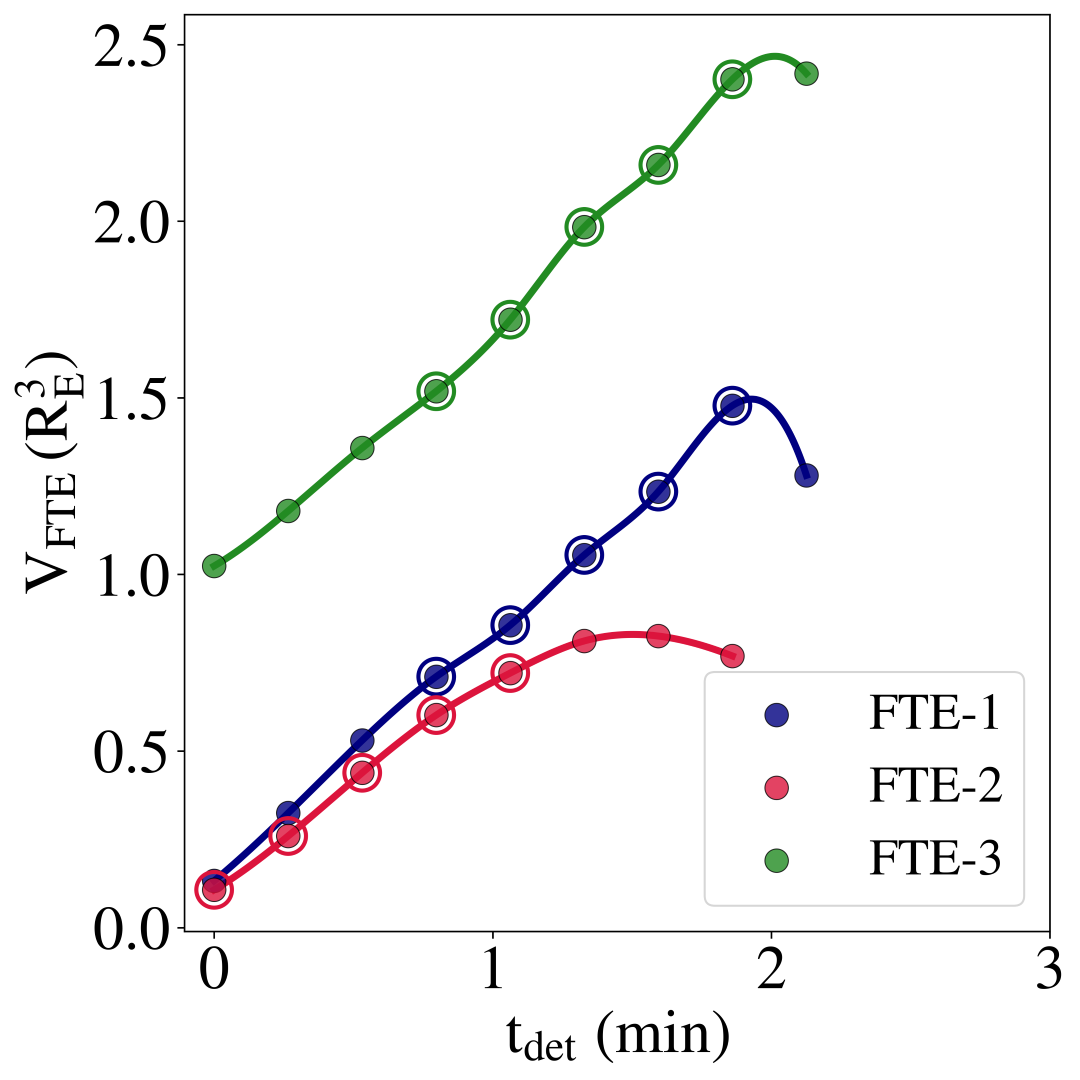}
    \caption{Plot shows the time evolution of volume of the selected FTEs. The abscissa shows $\rm t_{det}$ which represents the time passed after the FTE was first detected in the domain and the ordinate denotes the FTE volumes in units of $\rm R_E^3$. The encircled points have been selected to plot Figure \ref{fig:PV_diagram}.}
    \label{fig:vol_vs_time}
\end{figure}

As stated before, due to constraints associated with spacecraft trajectories, it is generally difficult to extract precise information about the azimuthal extent of the FTEs which makes the inference of the total FTE volume from observational data extremely challenging. From a numerical perspective, however, the volumetric detection paves the way for such a study. Figure \ref{fig:vol_vs_time} shows the time evolution of the volume of the selected FTEs in our simulation. The volumes are given in units of $\rm R_E^3$. For comparative purposes, the abscissa denotes time in $\rm t_{det}$ which stands for the `time of first detection'. This corresponds to the time elapsed after the FTE was first volumetrically detected in the simulation. For each FTE, $\rm t_{det}= 0$ therefore, corresponds to the global time given in the second column of Table \ref{Table:FTE}.

All the three selected FTEs initially show an increasing profile in their volume over time. The scatter points in the plot represent the exact values obtained from the analysis which is then over plotted with cubic splines to highlight the trend. The volume grows, stabilises and eventually develops a tendency to decrease when the FTEs are farther away from the subsolar point. Of the three selected FTEs, the FTE-3 grows up to the largest size with a final volume of 2.42 $\rm R_E^3$ and the FTE-2 is the one with the smallest maximum size of 0.82 $\rm R_E^3$. FTE-1 attains a moderate maximum volume of 1.47 $\rm R_E^3$. We note here that conventional \textit{in-situ} satellite observations measure the flux-rope `size' as an equivalent diameter of a circular FTE cross-section traversed by the probe. This however, can lead to significant variations in the modelled radius depending on the relative trajectory of the FTE and the probe. For example, we find that a cross section of FTE-1 across the mid-point of its azimuthal extent yields a nearly elliptical slice. For a probe passing perfectly through the FTE core (impact parameter= 0), there exists two extreme possibilities. The FTE size measured by the probe along the longest path across the cross-section gives a diameter of 1.2 $R_E$ whereas the shortest path detects a diameter of 0.58 $R_E$. Size inferences from \textit{in-situ} observations are prone to such large variations and multi-point observations are therefore essential as they rely on more sophisticated reconstruction techniques to infer the FTE sizes, albeit at a much higher computational cost \citep{Sonnerup_2004,Sonnerup_2006,Trenchi_2016}.

A salient feature of the temporal evolution of FTE volumes is that the volumes do not increase indefinitely with time. This is relevant from the perspective of the size distribution of the FTEs observed at the Earth's magnetosphere. All observations consistently agree that the size distribution of the FTEs falls off exponentially with increasing FTE size indicating that extremely large FTEs are less probable. In addition to the reason that the location of observation (near the subsolar point or further away) can have an inherent bias on the FTE size distribution, the fact that the FTE growths are stunted beyond a certain point may be a reasonable cause for this distribution.

We now examine in more detail, the primary reason for the growth of the FTEs. \citet{AkhavanTafti_2018} have summarised within a magnetospheric context that there can be three primary reasons for FTE growth. (a) The subsolar magnetosheath has a larger thermal pressure as compared to the flanks and an FTE can grow via adiabatic expansion when it travels away from the subsolar region. (b) The continuous reconnection process along the length of the evolving FTE can feed in hot plasma into the flux rope thereby increasing its size. (c) Analogous to the scenario of plasmoid merging in 2D, multiple smaller flux ropes may coalesce together in a self-similar fashion giving rise to a larger flux rope. The process of coalescence has additional complexities associated with it wherein, the angle between the axial components of the two flux ropes influence the merging process. Previous observations on planetary magnetospheres have indeed confirmed that either of the three processes may serve as the dominant cause of FTE growth \citep{Jasinski_2016, AkhavanTafti_2019}. In order to have a robust inference of the mechanism of FTE growth, it is preferable to study the individual FTE structures as they evolve. Presently, due to constraints associated with \textit{in-situ} observations, it is only feasible to study the temporal evolution of the thermodynamic characteristics of an individual FTE from a numerical perspective. To probe into this, we examine the P-V (Pressure-Volume) characteristics of the three FTEs featuring the dependence of the volume averaged pressure of the FTEs on their corresponding volumes. For this purpose, we select the regions of steepest descent within the P-V curves for each FTE. These sets of selected points for the corresponding FTEs are highlighted as encircled points in Figure \ref{fig:vol_vs_time}.

\begin{figure}
    \centering
    \includegraphics[width= 0.9\columnwidth]{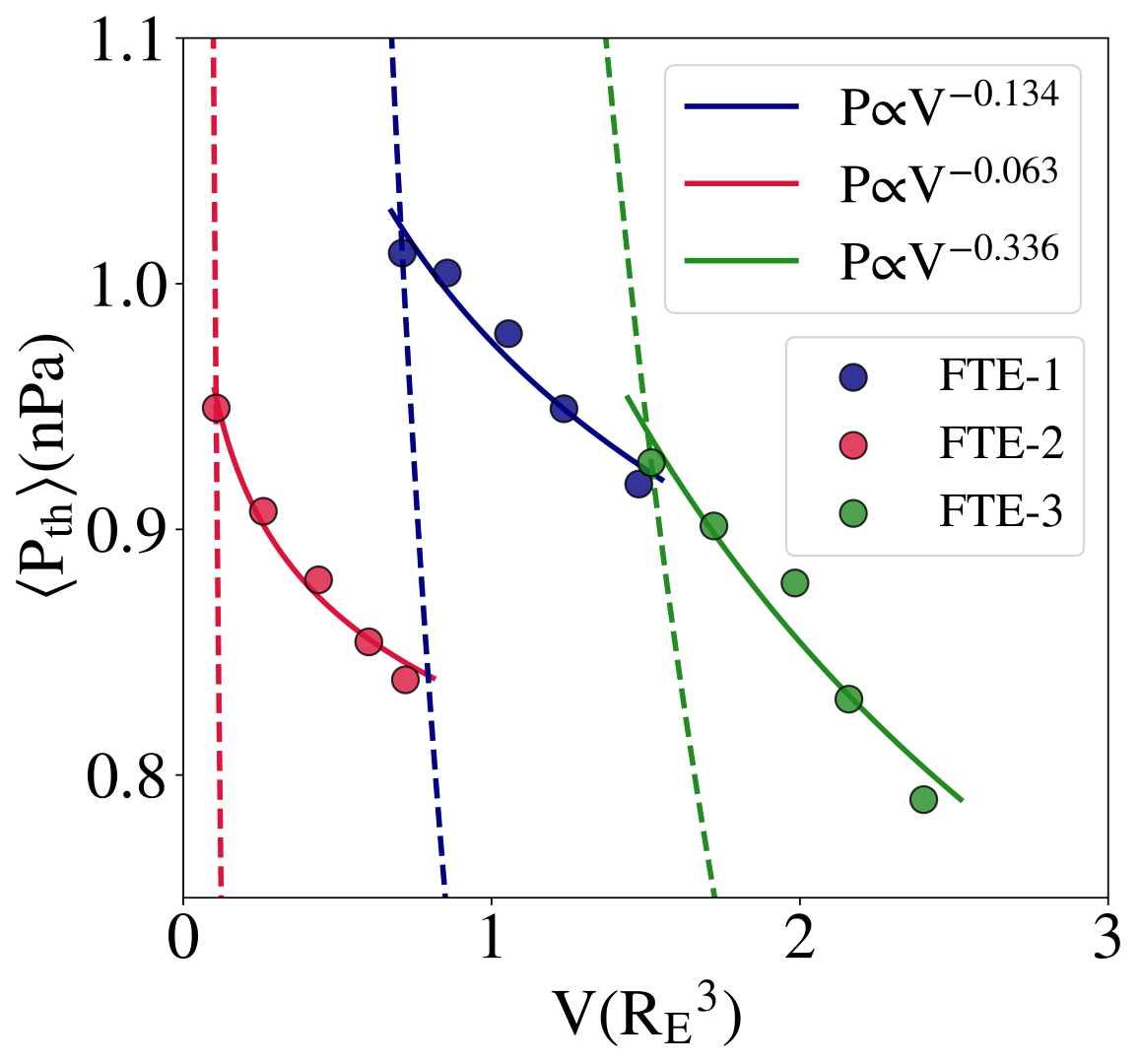}
    \caption{A plot of the P-V characteristic curves of the three FTEs. The plotted points are correspond to the ones encircled in Figure \ref{fig:vol_vs_time}. These sets of points are chosen as they exhibit the steepest descent in the P-V diagram. The correspondingly colored solid lines represent power-law fits to the data points. The correspondingly colored dashed lines represent the adiabatic ($\rm P \propto V^{-1.67}$) profiles passing through the first data point of each FTE.}
    \label{fig:PV_diagram}
\end{figure}

Of the three reasons of FTE growth cited above, in the case of FTEs 1, 2 and 3, island coalescence is not the cause of the growth of FTE volume as such a mechanism would have clear signatures of merger in the temporal profile of the $\rm B_{N}$ component of the magnetic field. As such signatures were not seen for the selected FTEs, the probable causes therefore lean towards either continuous reconnection, adiabatic expansion or a combination of the two processes. For adiabatic expansion to be the principal cause of FTE growth, one expects the pressure to fall according to the equation $\rm P \propto V^{-\gamma}$. For a monoatomic ideal gas, $\rm \gamma= 5/3$, therefore a sharp decrease in pressure is expected as the FTEs expand. 

The scatter points in figure \ref{fig:PV_diagram} plots, for each FTE, the set of encircled points of figure \ref{fig:vol_vs_time} in a P-V diagram. The ordinate corresponds to the thermal pressure averaged over the FTE volume in nanopascals (nPa) and the abscissa represents the FTE volume in $\rm {R_E}^3$. The solid lines represent the corresponding power law fits. The FTEs 1, 2 and 3 fit profiles of $\rm P \propto V^{-0.134}$, $\rm P \propto V^{-0.063}$ and $\rm P \propto V^{-0.336}$  respectively. For reference, we have also included the adiabatic expansion profiles ($\rm P \propto V^{-1.67}$) corresponding to the three FTEs passing through the first data point of each FTE in figure \ref{fig:PV_diagram}. They are represented by the appropriately colored dashed lines. It is clear that the curves obtained from the simulation are not compatible with the expected $\rm P \propto V^{-1.67}$ profile by a substantial margin. Such a consistently weak decrease essentially means that any drop in the thermal pressure is being replenished rapidly as the FTE expands. This replenishment can be attributed to the hot plasma exhausts associated with the ongoing reconnection in the neighbouring X-lines entering the flux rope \citep{Drake_2014}. We therefore assert that the process of ongoing continuous reconnection is the dominant cause of FTE growth for the FTEs 1, 2 and 3. 

Converting our P-V relation to a pressure-diameter relation assuming $\rm V\propto diameter^2$ for a perfectly cylindrical flux rope of constant length yields a value of $\rm P \propto (diameter)^{-0.35}$, where the exponent is averaged over the three FTEs. The pressure-diameter relation is a dependence that can be inferred directly from observations. An ensemble study on pressure-diameter relations of a set of 55 FTEs observed at the Earth's magnetopause by \citet{AkhavanTafti_2019} have also led to a similar relation of $\rm P \propto (diameter)^{-0.24}$ based on which they have concluded that continuous reconnection can be the dominant cause of FTE size increase at the Earth's magnetosphere. A caveat of the pressure-diameter relation is the inherent assumption that the length of the flux rope remains constant throughout the evolution, however, we find from our simulation that this constraint is too strict and the length of the flux ropes generally increase as they evolve. Even so, our results are similar to that of \citet{AkhavanTafti_2019} and concur with their assertion that the dominant cause of FTE growth at the Earth's magnetopause, under generic conditions, is the process of continuous magnetic-reconnection. We also note here that the process of FTE growth by coalescence is indeed seen in certain FTEs from our simulation (not in FTEs 1, 2 and 3). However, we find that such FTEs show a rapid jump in the FTE volume over time in contrast to the relatively smooth profile as seen in figure \ref{fig:vol_vs_time}. We also emphasise a deduction here that for FTE-1 and FTE-3, the steepest descent of the P-V curves occur towards the later phases of the FTE evolution (see figure \ref{fig:vol_vs_time}). Prior to this, the drop in pressure was slower. This transition may be considered as an indication that as the FTEs evolve, the contribution of adiabatic expansion to the volume growth of the FTEs increase. Thus, under the condition that any active reconnection processes have subsided, towards the very late stages of FTE evolution, the FTEs will eventually drift towards an expansion which is purely adiabatic.

\subsection{Magnetic Properties}\label{sec:magnetic_prop}

The idea that FTEs are flux ropes formed due to multiple X-line reconnection at the Earth's magnetopause is widely agreed upon and is supported by multiple observations and simulations \citep{Paschmann_1982, Lv2016,Chen_2017, Sun_2019,Mejnertsen_2021}. In this section we analyse in detail, the magnetic properties within the selected FTEs. 
\subsubsection{Force-Free Nature of FTEs}
As stated in section \ref{sec:init_evol_char}, the FTEs formed in our simulation take the form of twisted helical magnetic flux ropes. Magnetospheric flux ropes are often modelled to be magnetically `force-free' structures. A completely force-free configuration mandates that all currents are field aligned (in addition to the pressure gradient being negligible), i.e; 
\begin{equation}\label{eqn:jcrossb_j=alphab}
\begin{aligned}
    |\textbf{J}\times\textbf{B}|= 0   \quad ; \quad  \textbf{J}=\alpha(\textbf{r}) \textbf{B}
\end{aligned}
\end{equation}
where $\alpha$ is a function of position (\textbf{r}). The special case of $\alpha$ having a constant value is referred to as the linear force-free flux-rope model which leads to the Lundquist solution to equation \ref{eqn:jcrossb_j=alphab} \citep{Lundquist_1950, Burlaga_1988}. Hereafter, the terms `linear force-free' and `constant-$\alpha$ force-free' has been used congruously with each other. The Lundquist solution to the above set of equations in cylindrical (r, $\phi$, z) coordinates is given by: 
\begin{equation}\label{eqn:Lundquist-sol}
\begin{aligned}
    B_r &= 0 \\
    B_{\phi} &= B_0 J_{1}(\alpha r)\\
    B_{z} &= B_0 J_{0}(\alpha r)
\end{aligned}
\end{equation}
where $B_0$ is the axial magnetic field component of the flux rope having a maxima at the flux rope center and $J_0$ and $J_1$ are the Bessel functions of the first kind. Such a linear force-free configuration is of particular interest as it represents a minimum energy configuration of twisted magnetic field lines, and therefore, are the expected final states of flux-rope evolution \citep{Woltjer_1958}. Owing to its simplicity, the constant-$\alpha$ force-free flux-rope model has been widely used to deduce the physical parameters of magnetospheric flux-ropes, such as FTE size, flux contents and the impact parameter of the spacecraft with respect to the flux-rope core.

The following segments of this section have two objectives. First, we examine the temporal evolution of the perpendicular currents within the FTE volumes in order to investigate the magnetic configuration of the FTEs 1, 2 and 3. This is to discern if the currents within FTEs drift towards a field-aligned configuration as has been widely regarded in observations. Thereafter, we proceed to further examine the validity of a linear force-free flux-rope model within the context of the selected FTEs.

From an observational perspective, flux ropes which are in their `old/mature' stages are generally treated as magnetically force-free. This is in view of the fact that such a condition is an evolutionary acquirement and not an inherent constraint on the formation mechanism. A mandate of the force-free assumption is that all currents must be field aligned, i.e, $\rm \lvert \textbf{J}\times\textbf{B} \rvert\sim 0$. With regard to the first objective, to study the evolution of perpendicular currents within an FTE, we define a quantity $\mathcal{F}$ which is mathematically given as:
\begin{equation}\label{eqn:F_J_crossBbyBsq}
    \mathcal{F} =  \biggl< \bigg\lvert \frac{\textbf{J}\times\textbf{B}}{B^2}\bigg\rvert \biggr>_{vol}
\end{equation}
where $<>_{vol}$ denotes volume average. The term $\rm |\textbf{J}\times\textbf{B}|$ is directly derived as the Lorentz force which is then normalised to the square of the field magnitude $\rm |B|^2$. An increasing field aligned current would naturally mean $\mathcal{F} \rightarrow 0$. We note here that this study does not pertain to the force-balance within FTEs but rather highlights the evolution of the non-field aligned currents within the structure.
\begin{figure}
    \centering
    \includegraphics[width= 0.9\columnwidth]{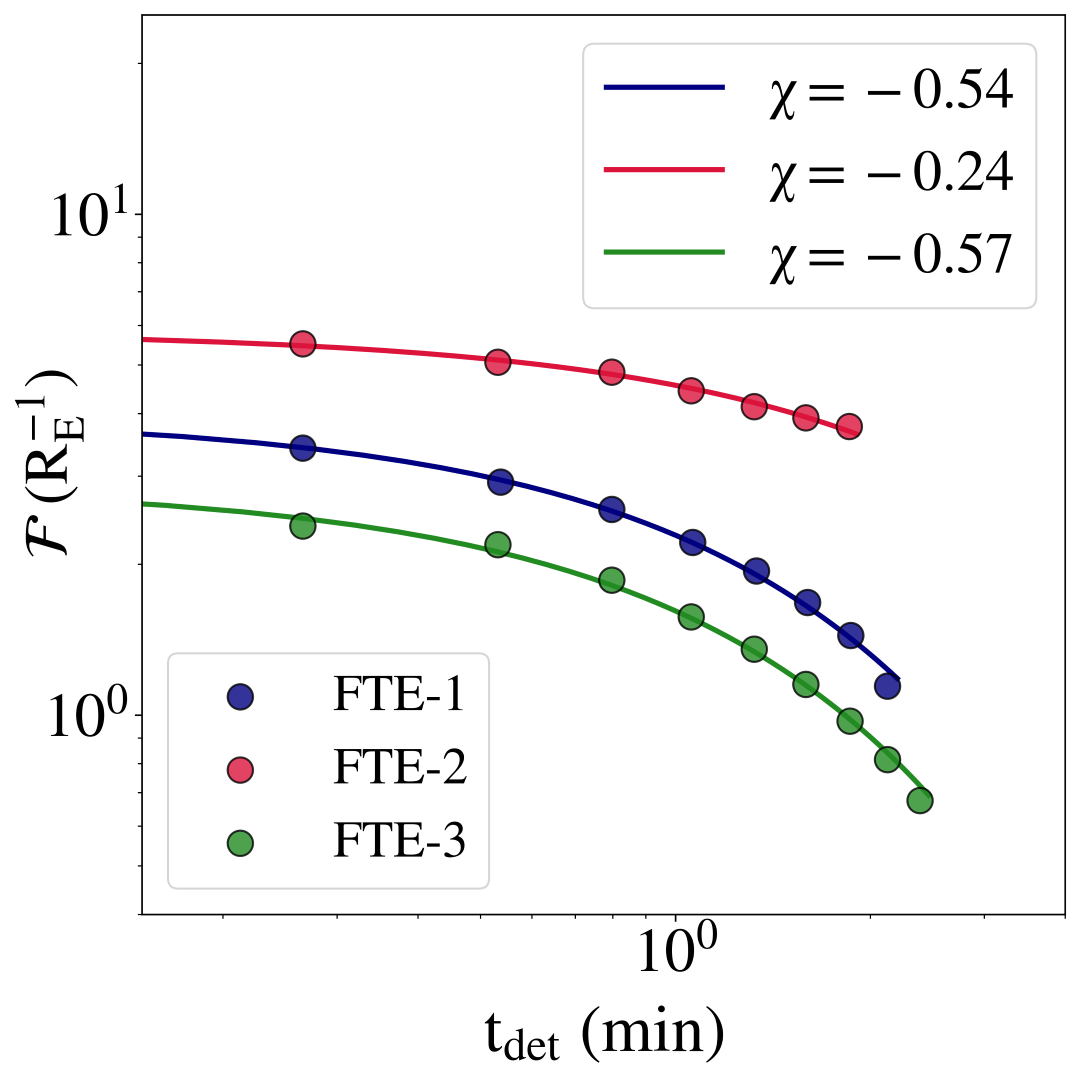}
    \caption{Plot shows the volume averaged value of $\mathcal{F}$ over time (see section \ref{sec:magnetic_prop} for details). The scatter points represent the values obtained from the simulations whereas the solid lines represent fitted exponential functions.}
    \label{fig:jcrossb}
\end{figure}

Figure \ref{fig:jcrossb} plots the volume averaged $\mathcal{F}$ of the individual FTEs with respect to time. For comparison, the time is again represented in terms of $\rm t_{det}$ as defined in section \ref{sec:Thermodynamic_props}. The colors denote the separate FTEs as per the legend. The quantity $\mathcal{F}$ is found to decay exponentially with time as seen by the exponential fits of the form $\mathcal{F}\propto e^{\chi t_{\rm det}}$. The exponential fits are denoted by the appropriately colored solid lines in figure \ref{fig:jcrossb}. This exponentially decaying behaviour is consistent across all the three selected flux ropes. The corresponding exponential functions are given by $\rm \mathcal{F}\propto e^{-0.54t_{\rm det}}$, $\rm \mathcal{F}\propto e^{-0.24t_{\rm det}}$ and $\rm \mathcal{F}\propto e^{-0.57t_{\rm det}}$ for FTEs 1, 2 and 3 respectively. This result is a demonstration to the widely accepted notion that old/mature magnetospheric FTEs have currents that are largely field-aligned ($\rm |\textbf{J}\times\textbf{B}|\sim 0$). Furthermore, looking at the trajectory of the FTEs in our simulation, one would expect that observations that are conducted further away from the generation region of FTEs would find flux-ropes that have more field aligned currents than any observations close to the generation region. 

Observations of magnetospheric FTEs have been extensively modelled using the linear (constant-$\alpha$) force-free flux-rope model \citep{Eastwood_2012,AkhavanTafti_2018, AkhavanTafti_2019} motivated by the analysis that ($\rm |\textbf{J}\times\textbf{B}|\sim 0$) within the FTE structure. Even though we find that the currents within the simulated flux-ropes rapidly evolve towards a field aligned form, that does not automatically guarantee that such a configuration will conform with a constant-$\alpha$ state. Here, we investigate the validity of a linear force-free model within the capacity of our simulations. Deriving from equation \ref{eqn:jcrossb_j=alphab}, the force-free parameter `$\alpha$' can be written as:
\begin{equation}\label{eqn:jdotbbybsq_alpha}
    \alpha= \frac{\textbf{J}\cdot \textbf{B}}{B^2}
\end{equation}
As an investigation towards the validity of a constant-$\alpha$ approximation for a magnetospheric flux rope, we probe into the distribution of the $\alpha$ values calculated at all the grid cells constituting the FTE volume. 

\begin{figure}
    \includegraphics[width= 1.0\columnwidth]{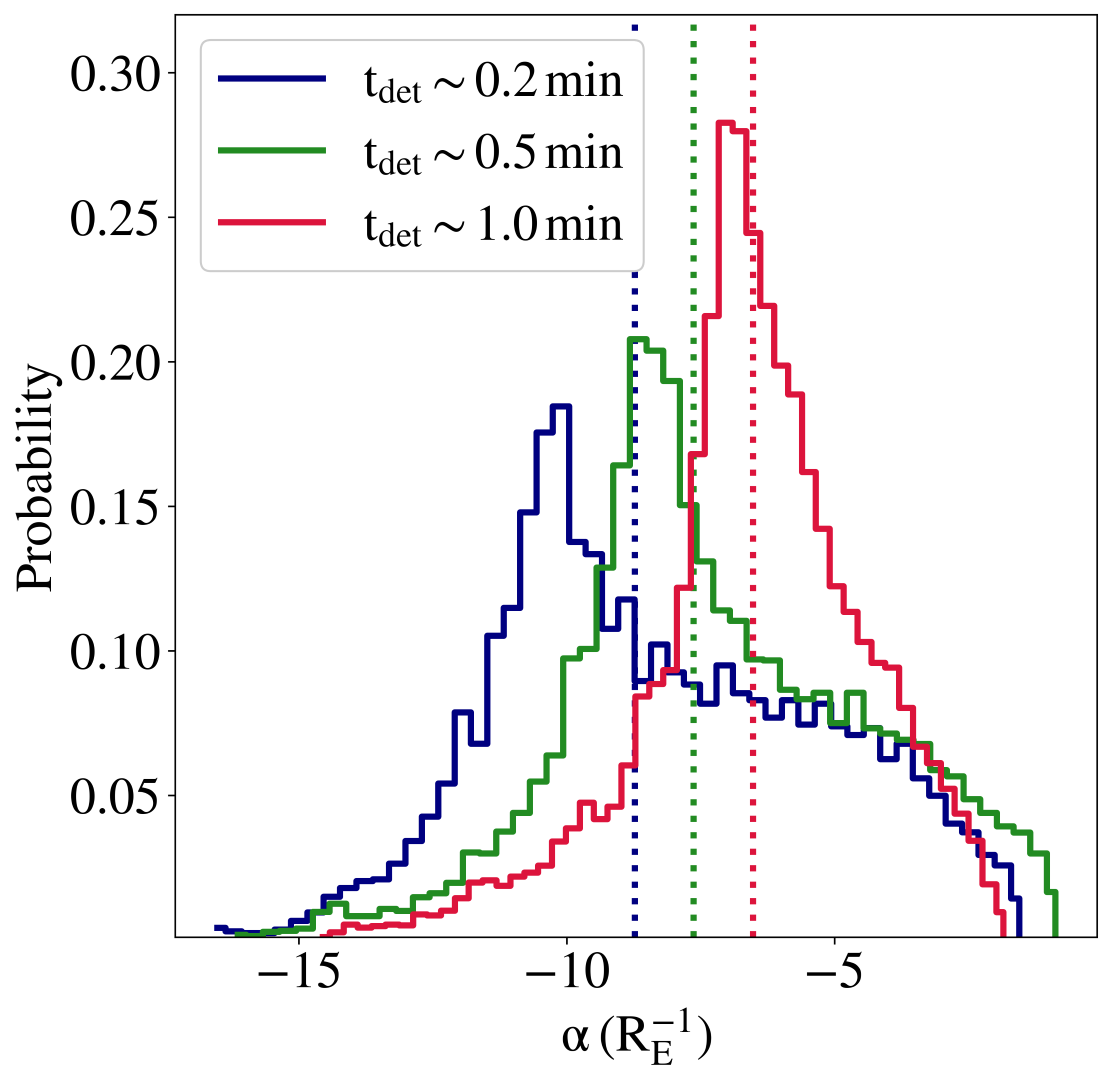}
    \caption{Plot shows the normalised histograms of the distribution of $\alpha$ within the volume enclosed by FTE-1 at different time snapshots as per the legend. The correspondingly colored dotted vertical lines represent the median value of the distribution at that instant.}
    \label{fig:hist_at_diff_times}
\end{figure}

For an ideal, purely cylindrical,  constant-$\alpha$ force-free flux rope, one would obtain a single value of $\alpha$ throughout the FTE volume, i.e, a histogram of such a distribution of $\alpha$ would have a single peak at a constant value. The histogram obtained from our simulations, however, exhibits a significant spread of $\alpha$ values for the very early phases of the temporal evolution of each flux rope. This can readily be seen from the blue curve in figure \ref{fig:hist_at_diff_times} which is a histogram of the values of $\alpha$ within FTE-1 at $\rm t_{det}\sim0.2 min$. The abscissa in figure \ref{fig:hist_at_diff_times} represents the $\alpha$ values whereas the ordinate shows the fraction of grid cells within the FTE structure corresponding to that value (bin) of $\alpha$.
This is expected, as the reconnection process leading to the formation of FTEs does not have any associated constraint to assume a constant-$\alpha$ form in terms of the magnetic field configuration. The temporal evolution of the histogram, however, exhibits the formation of a skewed distribution with a growing peak. This is evident from the three histograms depicted in figure \ref{fig:hist_at_diff_times}. The blue, green and red curves have been plotted in time snapshots corresponding to $\rm t_{det}\sim0.2\, min$, $\rm t_{det}\sim0.5\, min$ and $\rm t_{det}\sim1.0\, min$ of FTE-1. A closer look reveals that the peak becomes narrower and more significant as time progresses. This essentially means that with an increase in time, an increasing number of grid cells are converging towards a constant value of $\alpha$ throughout the entire volume of the FTE. This trend is seen for all three FTEs under consideration. The correspondingly colored dotted vertical lines in figure \ref{fig:hist_at_diff_times} represent the median value of each of the distributions. The median of the distribution is also seen to shift towards smaller magnitudes as time progresses.  This drift of the median value is anticipated because, mathematically, $\alpha$ behaves as an inverse length scale and a linear force-free configuration with a lower magnitude of $\alpha$ would have a larger diameter. As the FTEs within our simulation grow in size, the larger cross sectional area dictates that a force-free configuration for a similar size will have a lower magnitude of $\alpha$.
\begin{figure}
    \centering
    \includegraphics[width= 0.9\columnwidth]{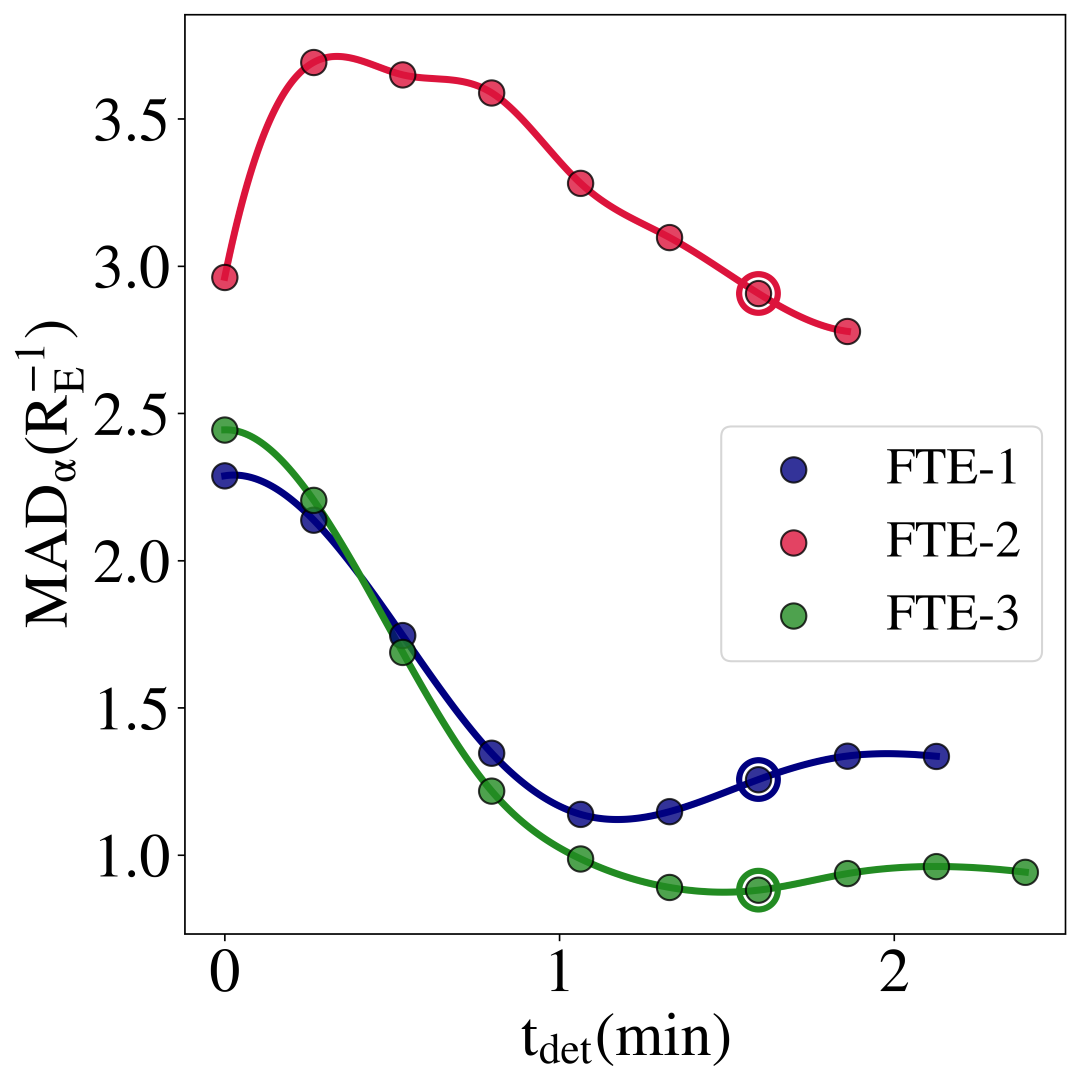}
    \caption{Plot shows the variation of the median absolute deviation of $\alpha$ with time. The scatter points represent the values obtained from the simulations whereas the solid lines are cubic interpolations representative of the trend. The encircled point on the FTE-1 curve corresponds to the time when Figure \ref{fig:constant_alpha_FFFR} was plotted whereas the encircled points for all the three FTEs combined were used for flux estimates summarized in table \ref{Table:Flux_Calc}}
    \label{fig:alpha_vs_time}
\end{figure}

We quantify the transformation of the spread of $\alpha$ throughout the FTE volume in figure \ref{fig:alpha_vs_time}, which shows the time evolution of the median absolute deviation on $\alpha$ ($\rm MAD_{\rm \alpha}$) as obtained from the distribution. The $\rm MAD_{\rm \alpha}$ for all three FTEs show a general decreasing trend as the FTE evolves, which characteristically means that the entire volume of the FTEs gravitate towards a constant-$\alpha$ configuration with time. However, towards the later stages of the evolution (after $\rm t_{det}\sim 1.0 min$), the $\rm MAD_{\rm \alpha}$ for FTE-1 and FTE-3 stops decreasing and stabilizes around a relatively low value. FTE-2 is seen to have an initial increase in the $\rm MAD_{\rm \alpha}$ during the very early stage of evolution ($\rm t_{det} \lessapprox 0.5 min$) followed by a continuously decreasing profile thereafter. The trend, however, indicates that given enough time, the $\rm MAD_{\rm \alpha}$  for FTE-2 would eventually stabilize akin to FTE-1 and FTE-3. We have verified that other statistical metrics of spread, e.g, mean absolute deviation, standard deviation etc., show a very similar behaviour for all the three FTEs. Succinctly stated, all three curves unanimously indicate that the FTEs do tend to drift rapidly towards a constant-$\alpha$ force-free state. Nevertheless, there always exists a finite deviation from a constant alpha value which inhibits the FTEs from attaining a purely linear force-free state.

\begin{figure*}
    \centering
    \includegraphics[width= \linewidth]{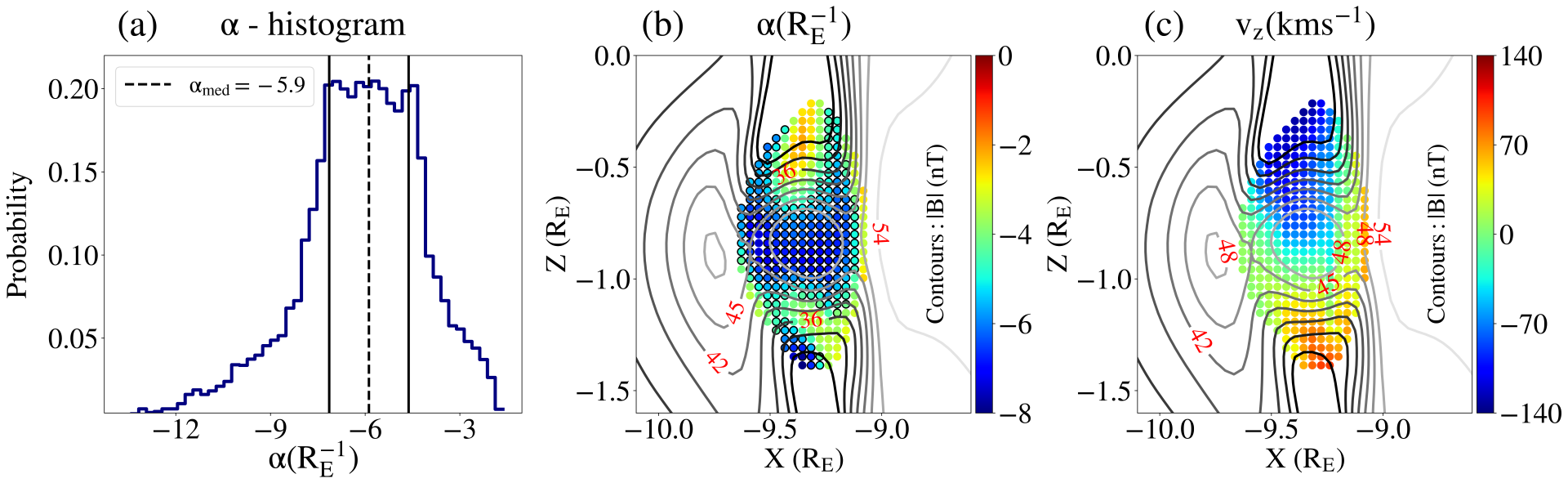}
    \caption{Panel (a) shows a normalised histogram of $\alpha$ taken over all the grid cells constituting the volume of FTE-1 at $\rm t_{det} = 1.6$min (corresponding to the encircled point on the FTE-1 curve in Figure \ref{fig:alpha_vs_time}). The dashed vertical line shows the median of the distribution and the two solid lines represent a span of unit median absolute deviation (MAD) on either side. The colored points in panel (b) represents the $\alpha$ on a slice of FTE-1 at y= 4.5 $\rm R_E$ whereas points encircled in black denote that these points lie within unit MAD of the median. Colored background points in panel (c) correspond to $\rm v_z$ on the same slice. The gray-scale contours in panels (b) and (c) correspond to the magnitude of the total magnetic field $\rm |\textbf{B}|$ where lighter shades represent higher values. The numbers over the contour lines represent the value of $\rm |\textbf{B}|$ in nanoteslas (nT).}
    \label{fig:constant_alpha_FFFR}
\end{figure*}

To discern the reason for the stagnation of $\rm MAD_{\rm \alpha}$, we investigate the distribution of $\alpha$ within FTE-1 during a time when the $\rm MAD_{\rm \alpha}$ value has stabilised. This temporal point is denoted by the blue encircled point on the curve corresponding to FTE-1 in figure \ref{fig:alpha_vs_time}. Panel (a) in figure \ref{fig:constant_alpha_FFFR} shows a normalized histogram of the distribution of $\alpha$ within the FTE-1 volume. As mentioned earlier, the distribution profile is skewed with a significant peak close to a median value. The median value $\rm \alpha_{med}$ was calculated to be -5.9 $\rm R_{E}^{-1}$ which is denoted by the dashed vertical line. The two solid lines on both sides denote the $\rm \pm 1 MAD_{\alpha}$ from the median value. As seen, a substantial portion of the peak lies within $\rm \pm 1 MAD_{\alpha}$ of the median value.
Panel (b) of figure \ref{fig:constant_alpha_FFFR} shows the distribution of the value of $\alpha$ within a cross section of FTE-1. The cross section is a slice at y= 4.5 $\rm R_E$ which is approximately halfway along the azimuthal extent of the FTE. Each point constituting the slice corresponds to the $\alpha$ value of the grid cell at that location. The encircled points within the cross section (points encircled in black) highlight the set of locations having an $\alpha$ value that lies within $\rm \pm 1 MAD_{\alpha}$ of the median $\alpha$. We see that the value of ${\alpha}$ for a majority of the points constituting the flux rope core falls within $\rm \pm 1 MAD_{\alpha}$ of the median value. In particular, nearly 60 \% of the FTE cross section falls within the $\rm \pm 1 MAD_{\alpha}$ bounds. The top and the bottom peripheral regions of the FTE cross section, however, have ${\alpha}$ values that lie outside these bounds.

To probe further into this, we look into the distribution of the z-component of velocity ($\rm v_z$) within this cross section. It is seen that there exists a significant counterstreaming flow of $\rm v_z$ at precisely the  same regions where the values of ${\alpha}$ in panel (b) deviate outside the $\rm \pm 1 MAD_{\alpha}$ bounds. The counterstreaming plasma inside the FTEs are in fact, the reconnection exhausts produced by the X-line pair outside the FTE boundary that feed in plasma into the FTE volume. We therefore deduce that the exhausts produced by the active X-lines on the top and bottom edges of the FTEs feed in newly reconnected flux that do not conform to the pre-existing peaked distribution of the ${\alpha}$ value. Therefore, as long as there is an ongoing continuous reconnection process, the FTEs can never, in principle, attain a purely constant-${\alpha}$ state due to the persistent flux injection. The nearly steady state in the temporal evolution of $\rm MAD_{\alpha}$ for the case of FTE-1 and FTE-3 as seen in figure \ref{fig:alpha_vs_time} is thus a sustained conflict between the flux rope tending towards a constant-${\alpha}$ form and the newly-reconnected flux disrupting this state. The core region of the flux rope, is however, found to be resilient to such perturbations and remains in a nearly constant-${\alpha}$ state up to a reasonable extent. The plot of $\rm v_z$ (panel (c) of figure \ref{fig:constant_alpha_FFFR}) is also an attestation to our inference in section \ref{sec:Thermodynamic_props} that the process of continuous reconnection feeds in hot plasma into the FTEs during their evolution.

\subsubsection{Flux Content Estimates}\label{sec:Flux_content}
The total flux content of FTEs is an important metric as it can provide an estimate of the contribution of such phenomena in the transfer of magnetic flux within the Dungey cycle \citep{Fear_2019_merc}. Within the linear force-free flux rope model, \citet{Eastwood_2012} presented an analytical estimate of the flux content  of a perfectly cylindrical flux-rope as:
\begin{equation}\label{eqn:analytical_flux}
    \Phi_{ana} = \left[\frac{ 2B_0 A_{fr} J_1 (2.40842)}{2.40842}\right] 
\end{equation}
where $B_0$ is the axial field strength at the flux-rope core, $A_{fr}$ is the cross sectional area of the flux-rope and $J_1$ is the Bessel function of first kind and first order. The core field strength $B_0$ and the area of the flux-rope $A_{fr}$  are determined from observations by the linear force-free model fits which are then fed into equation \ref{eqn:analytical_flux} to estimate the flux \citep{Eastwood_2012,Jasinski_2016,AkhavanTafti_2018,AkhavanTafti_2019}.

The following exercise is to compare the flux transported by our simulated flux-ropes to the flux enclosed by the analytical estimates of an analogous, perfectly cylindrical, linear force-free flux rope as calculated from equation \ref{eqn:analytical_flux}. Within our simulation, we estimate the flux content of the detected FTEs using the following procedure. Firstly, we span the entire azimuthal extent of the FTE to find the y-slice that shows the strongest value of the $B_y$ (axial) component. Thereafter, we consider a slice of the FTE at this particular value of the y-coordinate and calculate the flux content ($\Phi_{sim}$) using the expression of equation \ref{eqn:simulated_flux}, where the summation ($\Sigma$) is taken over all the points constituting the flux rope cross section.
\begin{equation}\label{eqn:simulated_flux}
    \Phi_{sim} = \Sigma (|B_y| .dA)  
\end{equation}
We then examine the cross section of the sliced FTEs, and locate the point where the axial component ($B_y$) is maximum. A region spanned by 3$\times$3 grid cells around this point is then selected and the median $B_y$ within that region is estimated as the core axial field strength ($B_0$) of the flux rope. This value, in addition to the cross sectional area of the flux rope (calculated according to the number of grid cells in the flux rope cross-section), is then used in equation \ref{eqn:analytical_flux} to estimate the flux content of an equivalent, purely linear, force-free flux rope. The flux value thus obtained is named as $\rm \Phi_{ana}$ and is a representative of the flux obtained for a satellite travelling through the FTE core, assuming that the cross sectional area is determined accurately. This process is repeated for all three FTEs at a time denoted by the encircled points in figure \ref{fig:alpha_vs_time}, i.e., during the significantly later stages of FTE evolution when the $\rm MAD_{\alpha}$ has stagnated.

Table \ref{Table:Flux_Calc} summarizes the results of the flux calculations using the aforementioned procedures. The first and second columns represent the names of the FTEs and the global time when the flux calculations were carried out respectively. The third and the fourth columns named `$\rm y_{beg}$' and `$\rm y_{end}$' represent the azimuthal extent of the corresponding FTEs in the domain. The column named `$\rm y_{slice}$' represents the location where the maximum value of $B_y$ was obtained for each FTE and therefore, corresponds to the coordinate where the FTE was sliced for the flux calculations. `$\rm B_{axial}$' quantifies this core magnetic field strength obtained for each FTE. $\rm \Phi_{sim}$ represents the flux values obtained from the simulations using equation   \ref{eqn:simulated_flux} and  $\rm \Phi_{ana}$ denotes the values obtained using equation \ref{eqn:analytical_flux} (in mega-Webers). The last column shows the difference in the two flux values in terms of the percentage error on $\rm \Phi_{sim}$.

The value of $\rm \Phi_{sim}$ for FTE-1 was calculated to be 0.52 MWb,  whereas the corresponding $\rm \Phi_{ana}$ had a value of 0.43 MWb. We note that FTE-1 has a moderate axial field strength and the total flux content was found to be nearly consistent between the simulated and analytical values. The flux content of FTE-2 was lower than FTE-1 where the corresponding  $\rm \Phi_{sim}$ and $\rm \Phi_{ana}$ was obtained to be 0.33 MWb and 0.30 MWb respectively. The errors in flux calculated by the analytical model of FTEs 1 and 2 were $\sim$17\% and $\sim$9\% respectively. The flux obtained for FTE-3 was $\rm \Phi_{sim}=$ 1.52 MWb, however, the analytical estimate was found to be $\rm \Phi_{ana}$=0.92 MWb. The total flux content of all the three FTEs, therefore,  were underestimated under the linear force-free assumption with FTE-3 exhibiting a significant underestimation of $\sim$40\%. We re-emphasise here that the flux obtained from the analytical model ($\rm \Phi_{ana}$) for a typical observation is in a scenario where it is assumed that the core magnetic field strength has been estimated correctly. 

We examine the distribution of $B_y$ (axial component contributing to the flux for the simulated FTEs) over all the points spanning the FTE-3 cross section as a means to explore the reason for such a significant underestimation. First, we analytically construct a perfectly cylindrical ideal flux rope which is in a constant-$\alpha$ force-free state. The magnetic field components given by equation \ref{eqn:Lundquist-sol} is therefore, a solution to such a flux rope. We then extract a random sample of points that uniformly span the circular cross section of the ideal flux rope and calculate the axial component ($B_z$ in equation \ref{eqn:Lundquist-sol}) at each sample location. The uniform sampling is done to imitate a Cartesian grid spanning the cross section. Even though the value of $\alpha$ used for this purpose is inconsequential, we have chosen a value which represents the median of the distribution of $\alpha$ obtained using equation \ref{eqn:jdotbbybsq_alpha} throughout all grid cells spanning the FTE-3 volume. As the axial component is the main contribution to the total flux content, the distribution of the axial component of these uniformly spaced samples within the ideal flux rope cross section is compared to the distribution of the axial component of all the points spanning the cross section of FTE-3 in our simulation.
\begin{figure*}
    \centering
    \includegraphics[width= 0.8\linewidth]{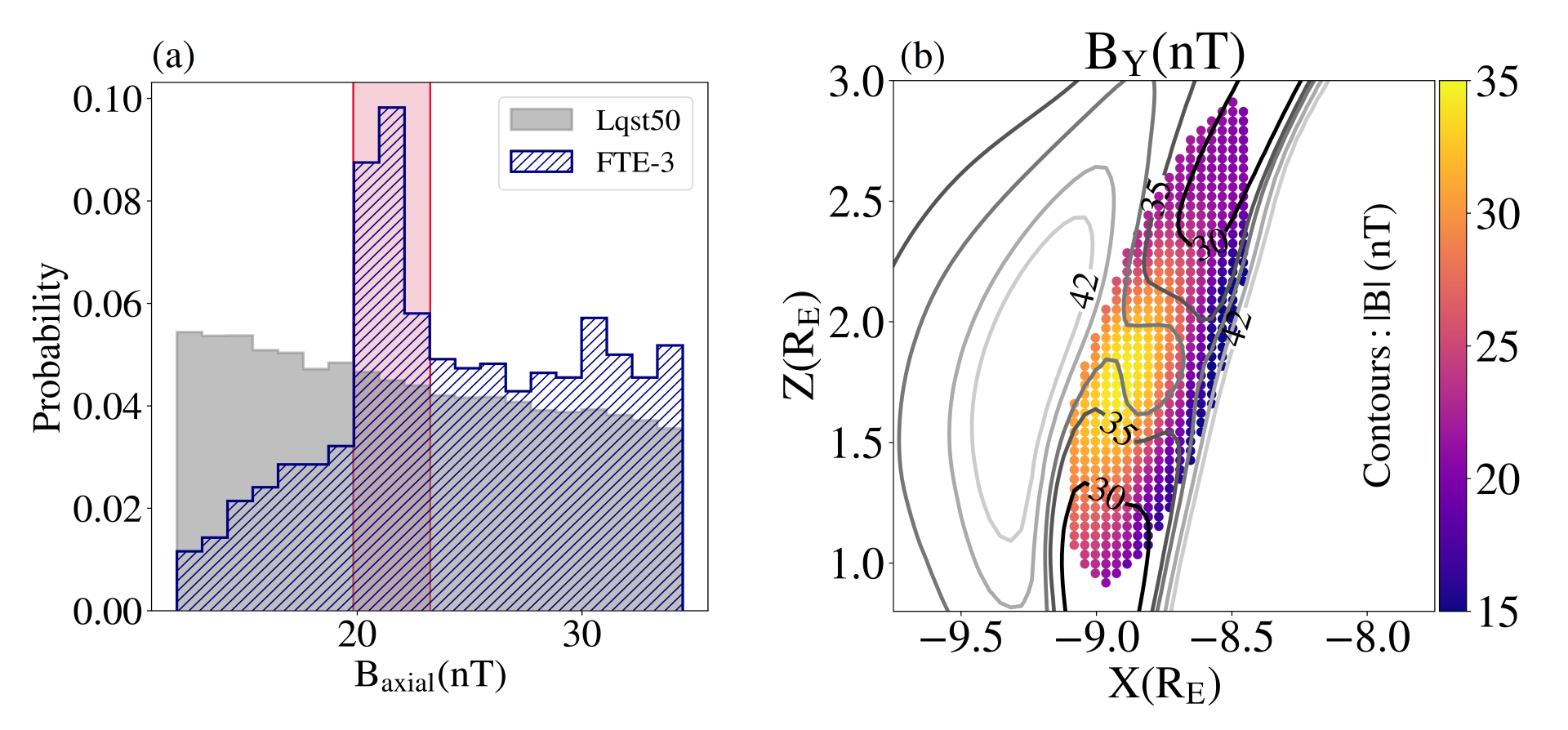}
    \caption{Panel (a) shows a plot of the normalised histogram of the axial components of the magnetic field obtained from the Lundquist solution (denoted as `Lqst50') i.e, equation \ref{eqn:Lundquist-sol} and from our simulations (denoted as `FTE-3'). See section \ref{sec:Flux_content} for details. Panel (b) shows an X-Z slice of the $\rm B_Y$ component within the FTE-3 cross section at $\rm y\sim4.8 R_E$. The gray-scale contours in panel (b) correspond to the magnitude of the total magnetic field $\rm |\textbf{B}|$ where lighter shades represent higher values. The numbers over the contour lines represent the value of $\rm |\textbf{B}|$ in nanoteslas (nT).}
    \label{fig:axial_component_histogram}
\end{figure*}
Panel (a) of figure \ref{fig:axial_component_histogram} shows the normalized histogram for both, the Lundquist solution (denoted by the grey shaded histogram) marked as `Lqst50' and the samples from the cross section of FTE-3 (denoted by the blue hatched histogram). There are two regions where the two histograms significantly differ, (a) the low axial field strength region (left portion of the histogram) and (b) the intermediate axial field strength region (near $\sim$21 nT). The differences between the two distributions at lower axial field strengths would have a non-trivial impact on the flux measurements, but the total flux content will be less susceptible to such types of deviations owing to the smaller field magnitude. The significant deviation denoted by the red shaded region in panel (a) of figure \ref{fig:axial_component_histogram} at intermediate axial field strengths, however, can have a crucial impact on the flux estimation due to the relatively higher magnitude as well as a much higher number of such points. A plot of the FTE-3 cross section at $\rm y\sim4.8 R_E$ is shown in panel (b) of figure \ref{fig:axial_component_histogram} for reference. We have verified that the points that lie within the red shaded region ($\rm B_{Y}\sim$21 nT) in panel (a) of figure \ref{fig:axial_component_histogram} are located at the top and the bottom region of the cross section of FTE-3 depicted in the panel (b). This strongly indicates that this excess is a direct implication of the newly reconnected magnetic flux being injected into the FTE volume from the X-lines associated with the FTE. With an intermediate field strength of $\sim$21 nT, regions like these can significantly enhance the flux content of an FTE. The linear force-free flux rope model does not take such a continuous injection of flux into account and has a strictly decreasing axial component as one moves away from the flux rope center. In other words, as one moves towards the edges of the flux rope, flux injected due to the ongoing reconnection process does not allow the axial component to drop as much as predicted by the linear force-free flux rope model. Therefore, the flux estimation performed using this model might, in these cases, largely underestimate the flux content of an FTE. As a final exercise in this context, to estimate the magnitude of the difference in $\rm \Phi_{sim}$ and $\rm \Phi_{ana}$ that can be attributed solely to this intermediate flux region, we recalculate the flux content of FTE-3 after excluding the points that constitute the red shaded region in figure \ref{fig:axial_component_histogram}. We find that this reduces the percentage error by almost half and brings the difference to a rather modest $\sim$20\% with the newly estimated flux value to be $\rm \Phi_{sim}= 1.15 MWb$. However, within the context of our simulations, we believe that a comprehensive statistical study of a significantly higher number of simulated FTEs will be required for a robust quantification of the discrepancies in flux context estimates within the FTEs.

Finally, we present a quantification on the total closed and open dayside magnetospheric flux in our simulation. The closed flux has been estimated as follows. During a time when the magnetopause is devoid of any transient phenomena, we take a slice of the computational domain at z=0 within the bounds $\rm -15 R_E \leq y \leq 15 R_E$ and $\rm -15 R_E \leq x \leq 0 R_E$. The bounds for x ensure that the slice only contains the dayside magnetosphere ($\rm x\leq 0 \Rightarrow \, Dayside$). The slice thus obtained is a 2D surface from which a region $\rm \sqrt{(x^2 + y^2)} \leq 1 R_E$ is excluded. This is to ascertain that the contribution to the flux is only from the region between $\rm 1 R_E$ and the magnetopause boundary. As the solar wind is composed primarily of a southward-IMF, all points on this surface that have $\rm B_{Z} >\,0$ (Earth's dipolar field lines) are taken to be the grid cells that belong inside the magnetopause thus contributing to the total closed flux and all points with $\rm B_{Z} < \,0$ (solar-wind field lines) are considered to lie outside the magnetopause and thus do not contribute to the total closed dayside flux. Adapting the approach from \citet{Tafti_2020}, the total closed dayside magnetospheric flux content ($\rm \Phi_{MSP}$) is then estimated using the following expression
\begin{equation}
    \rm \Phi_{MSP}= \sum_{\substack{\rm r>1  R_E\\ \rm \forall\; B_{Z}\geq 0}} B_Z \cdot dS 
\end{equation}
    where `dS' is the area of each grid cell composing the 2D slice. The total closed dayside magnetospheric flux content thus obtained from our simulation has a value of $\rm \sim 3.5\; GWb$. Thereafter, within a box bounded by  $\rm -15 R_E \leq x \leq 0 R_E$, $\rm -30 R_E \leq y \leq 30 R_E$ and $\rm -15 R_E \leq z \leq 15 R_E$, we also calculate the total dayside open flux as follows. From all points constituting the surfaces of this bounding box, we trace the magnetic field lines to determine if they reach 1 $\rm R_E$. The magnetic field lines thus isolated are the open field lines that originate from the surface of the Earth and reach the domain boundaries. For all the field lines traced from the outer boundary to 1 $\rm R_{E}$, we calculate the associated flux $\rm (\Sigma \mathbf{B_{i}} \cdot d\mathbf{S})$ where `$\rm \Sigma$' denotes summation over all such field lines, `$\rm \mathbf{B_i}$' and `d\textbf{S}' are the magnetic field and the area of the grid cell at the field line origin respectively. The total open flux of the dayside magnetosphere is thus calculated to be  $\sim$ 0.35 GWb. The flux content of FTE-3 (FTE-3 had the maximum flux content) turns out to be of the order of 1$\%$ of this total dayside magnetospheric open flux. As an additional note, we also find that this flux content was nearly 100 times lower than that of one of the large FTEs observed by \citet{Tafti_2020}. However, this estimated flux content of FTE-3 is only after $\sim$1.6 min of FTE evolution and as the FTEs evolve further, their flux content may tend to increase. We additionally report an estimation of the voltage across FTE-3 due to the ongoing reconnection. For this, we evaluate the initial (at $\rm t_{det}\sim 0$) flux content of FTE-3 having a magnitude of $\sim$ 0.4 MWb. This flux difference over an evolutionary time period then quantifies a voltage ($\rm \Delta \Phi_{sim} / \Delta t_{det}$) of $\sim$ 11.7 kV across FTE-3 due to reconnection.
\begin{table*}
\centering
 \begin{tabular}{|c|c|c|c|c|c|c|c|c|}
 \hline\hline
 Name  & t (s) & $\rm y_{beg}(R_E)$ & $\rm y_{end}(R_E)$ & $\rm y_{slice}(R_E)$ & $\rm B_{axial} (nT)$ & $\rm \Phi_{sim}(MWb)$ & $\rm \Phi_{ana}(MWb)$ & $\rm Error(\%)$ \\ 
  \hline
   FTE-1 & 1467 & 2.63   & 6.77  &    4.5  & 49.01 & 0.52  &  0.43  &  17 \\
   FTE-2 & 2902 & 3.45   & 5.91  &    4.90  & 32.67 & 0.33  &  0.30  &  9 \\
   FTE-3 & 3428 & -7.79  &-3.45  &   -4.83  & 34.45 & 1.52  &  0.92  & 40  \\
\hline\hline
 \end{tabular}
\caption{A table summarizing various parameters used for the flux estimation of the FTEs as well as the values and the errors between the flux calculated by the analytical model (equation \ref{eqn:analytical_flux}) and the flux obtained from our simulations (equation \ref{eqn:simulated_flux}).}
\label{Table:Flux_Calc}
\end{table*}
\section{Discussions and Summary}\label{sec:Discussion_and_summary}
We have investigated the evolution of FTEs formed due to localised reconnection at the dayside magnetopause of an Earth like planet. High resolution MHD simulation of the magnetosphere with adaptive mesh refinement (AMR) has been performed in this regard. We then leverage an agglomerative hierarchical clustering algorithm to volumetrically detect these arbitrarily shaped, 3-dimensional FTE structures in our simulation domain. Three FTE volumes have been selected and extensively analysed in this study with regard to their thermodynamic and magnetic properties. 
Altogether, we now summarize the principal findings of this study in the following sections.

\subsection{FTE Growth}
First, the time evolution of the volume of the FTEs were investigated for the FTEs 1, 2 and 3. The trend of FTE volume growth was found to be monotonically increasing during the early phases of FTE evolution followed by a stagnation of the growth. This stagnation can be attributed to either the relaxation of field lines within the FTEs or due to a drag induced compression  as the FTEs encounter fast moving magnetosheath flows \citep{Eastwood_2012}. Thereafter, we investigate the principal cause for the growth of the FTE volumes. An inspection of the P-V curves in figure \ref{fig:PV_diagram}, show that the average pressure drops much slowly in contrast to the relation $\rm P \propto V^{-1.67}$ expected from adiabatic expansion. Thus, we conclude that the process of continuous reconnection at the neighbouring X-lines inject hot plasma into the FTE volume which does not let the pressure decrease as rapidly as is expected from an adiabatically expanding FTE. In the absence of FTE coalescence, the process of continuous reconnection is, therefore, found to be the dominant cause of FTE growth for all FTEs analysed in this study. Altogether, we find our results to be synonymous with observations by \citet{AkhavanTafti_2019} which arrived at a similar conclusion. 

\subsection{Force-free Nature of FTEs}
An investigation into the magnetic properties within the FTEs show that the Lorentz force within the structures rapidly decreases. This indicates a swift tendency of any currents within the FTE volumes toward being field aligned as the FTE evolves. As generally asserted from observations, the trend of $\rm \lvert \textbf{J} \times \textbf{B}\rvert\, \rightarrow 0$ indicates a drift towards a magnetically force-free configuration. We therefore infer that our FTEs show a similar trend and eventually, would attain this $\rm \lvert \textbf{J} \times \textbf{B}\rvert\, \sim 0$, or magnetically force-free state. We further investigate the validity of the constant-$\alpha$ force-free flux rope model within the selected FTEs. We find that during the initial stages of FTE evolution, there exists a distribution of the value of $\alpha$ throughout the FTE volume. It is seen that as the FTEs evolve, this distribution exhibits a peak around a certain value indicating the tendency of the FTEs to gravitate towards a constant-$\alpha$ force-free form throughout the evolution.  We nevertheless observe that the process of continuous magnetic reconnection injects magnetic flux into the FTEs which tend to divert them from a constant-$\alpha$ configuration.

\subsection{Flux content of FTEs}
Complemented with observations to infer the physical properties of a detected FTE, the constant-$\alpha$ force-free flux-rope model is often used to deduce the flux content of a detected FTE. We calculate the flux content of the FTEs 1, 2 and 3 and compare it with a corresponding analytical estimate derived from the linear force-free flux rope model by \citet{Eastwood_2012}. The analytical calculations are found to underestimate the flux content in all three cases as shown in table \ref{Table:Flux_Calc}. The constant-$\alpha$ force-free flux-rope model does not take the aforementioned continuous flux injection into account and therefore underestimates the flux content by a significant margin. Our findings lean towards the argument given by \citet{Fear_2017} that in-situ measurements, in certain cases, do tend to underestimate the flux content of the FTEs. We note here that the study by \citet{Fear_2017} incorporates additional tools such as ionospheric radar measurements to aid in flux calculations.

In essence, our study addresses the evolution of the transient FTEs at the dayside magnetopause using numerical simulations. The novelty includes the adaptation of volumetric detection techniques to isolate the irregular 3-D FTE structures and study the temporal evolution of their thermodynamic and magnetic characteristics. This paper aims to bridge the interpretations obtained from in-situ observations of such FTEs, generally conducted along an FTE cross section, with numerical simulations illustrating a full 3-dimensional perspective. The paper highlights the detection and analysis of a subset of simulated FTEs. However, we note here that a statistical perspective with a large sample of FTEs is essential in order to have a robust assertion pertaining to the challenging issues such as the volume distribution of FTEs and flux content estimates. A statistical study that addresses these issues is in progress and shall be the focus of a future article.\\

\textbf{Acknowledgements:} The authors thank the reviewers for their constructive comments that have helped to improve the manuscript. The authors would also like to thank Andrea Mignone for his insightful suggestions. A.P. is a research scholar at IIT Indore and would like to acknowledge the support provided by the institute. B.V. is thankful for the support provided by the RESPOND grant (ISRO/RES/2/436/21-22) given by Indian Space Research Organization to support this research. He also acknowledges the support for leading the Max Planck Partner Group at IIT Indore provided by the Max Planck Society. A.S. acknowledges funding from the French Programme National Soleil-Terre (PNST) and Programme National de Planétologie (PNP). All computations presented in this work have been carried out using the facilities provided at IIT Indore and the Max Planck Institute for Astronomy Cluster: ISAAC, which is a part of the Max Planck Computing and data Facility (MPCDF). This research made use of \texttt{astrodendro} (http://www.dendrograms.org/), an open source Python package to compute dendrograms of Astronomical data.
\bibliography{ms_AP}{}
\bibliographystyle{aasjournal}
\appendix
\section{{Volumetric FTE Detection Technique}}\label{ss:detection methodology}
This section elaborates on the agglomerative hierarchical structure finding algorithm that has been used to perform volumetric detection of the FTEs in our simulation. Such forms of hierarchical structure identifications were first used in the astrophysics community to study \textit{structure-trees} in the context of star formation by \citet{Houlahan_1992} in 2-dimensional data. Similar techniques have also been adapted and extensively used in additional domains of astrophysics, e.g., in the study of 3-dimensional molecular-line position-position-velocity data cubes by \citet{Rosolowsky_2008}, in studying Giant Molecular Clouds by \citet{Piperno_2020} and in characterizing accretion flows around a protostellar system \citep{Thieme_2022}. Even though the form of the input data used for such classifications is predominantly observational, from a purely algorithmic perspective, the detection could be performed for any generalised input data set. Here, for the first time to our knowledge, we have extended the application of such a structure finding algorithm to detect 3 dimensional volumes enclosed by FTEs in a global-MHD simulation. 

The output of the algorithm is in the form of a dendrogram which is an abstraction of the structure hierarchy in the input data represented as a structure tree. The structure tree is composed of smaller structures in the form of ``branches'' which can be further decomposed into smaller independent structures called the ``leaves". The python package \texttt{astrodendro}\footnote{http://www.dendrograms.org/} has been used for this purpose. An illustration of the methodology used to detect structures in \texttt{astrodendro} can be found in the associated documentation\footnote{https://dendrograms.readthedocs.io/en/stable/algorithm.html}, however, we briefly describe the core algorithm here for completeness.
\begin{figure*}
    \centering
    \includegraphics[width= \textwidth]{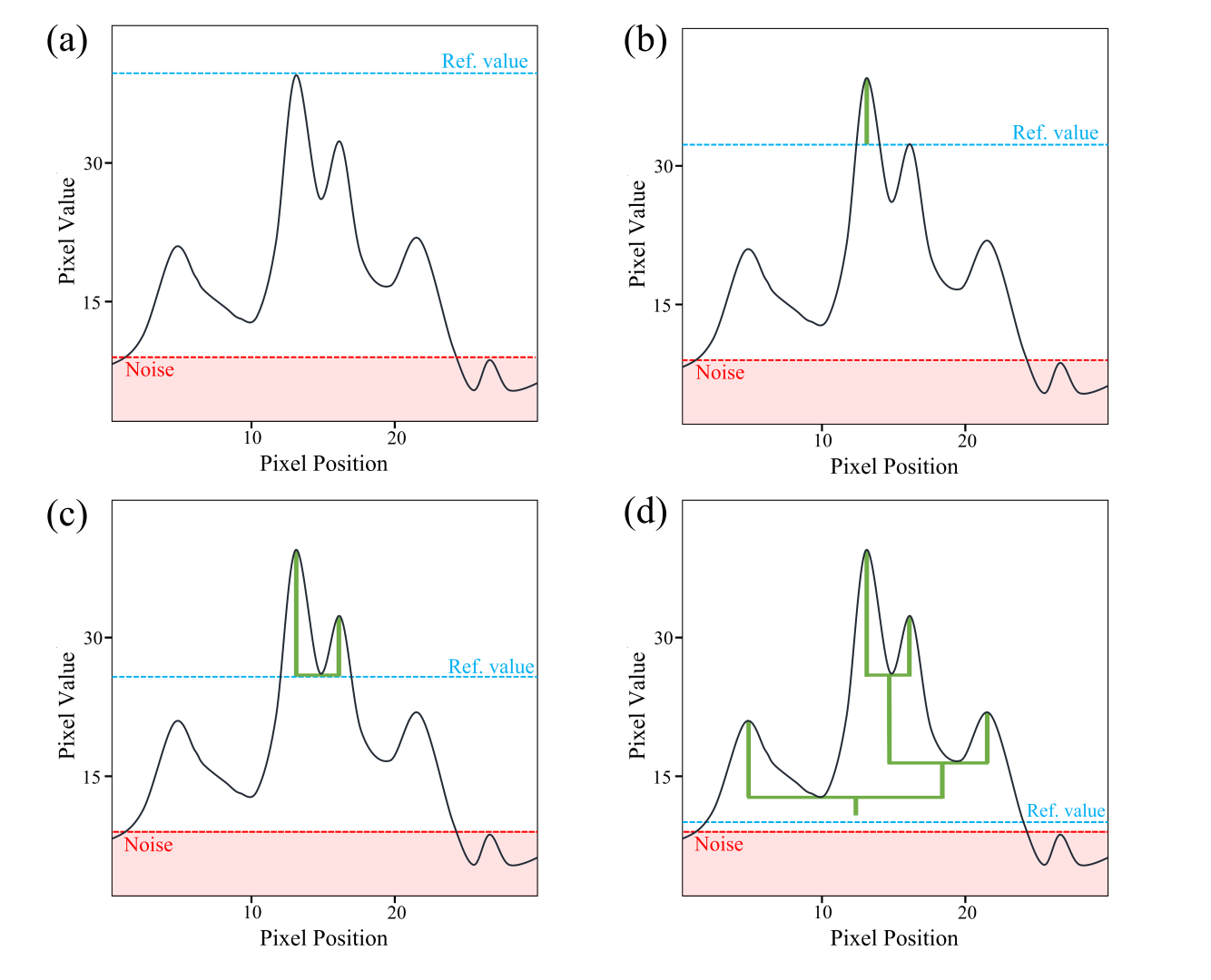}
    \caption{Panels (a), (b), (c) and (d) represent various stages of structure detection for a simplified one dimensional data set having a row of pixels with corresponding values associated with each pixel. The blue dashed line represents the reference value used by the algorithm which starts at the global highest pixel value and keeps decreasing. The red shaded portion in each of the panels denote the noise threshold below which the structures are not detected.}
    \label{fig:dendro_algo}
\end{figure*}

For simplification, let us consider a one dimensional data set having a row of pixels (or grid cells) and each pixel having an associated value. The black curve in panel (a) of figure \ref{fig:dendro_algo} represents one such data set. As seen, the profile contains multiple local maxima and minima. The ultimate goal of the algorithm is to construct a hierarchy of structures (structure tree) from this data set where each individual structure corresponds to the pixels surrounding a local maxima. One can also set a noise threshold on this data set which essentially means that all pixel values below this threshold will be considered as noise and shall not be incorporated in the hierarchical structure tree. This threshold level is denoted by the red shaded region in all panels of figure \ref{fig:dendro_algo}.

The algorithm starts at the global highest pixel value of the data set called the reference value. This has been marked with the dashed blue line in panel (a) of figure \ref{fig:dendro_algo}. This reference value now starts moving down from the highest peak which essentially marks the first structure in the hierarchy. This has been denoted by the vertical green line in panel (b) of figure \ref{fig:dendro_algo}. The reference value keeps moving down and encounters neighbouring pixels on both sides of the global maxima having lower values. For each pixel it encounters, the algorithm decides whether to join the pixel into an already existing structure or whether the pixel denotes the beginning of a new structure. A new structure is created only if the value of a pixel is higher than its immediate neighbours or in other words, if the pixel corresponds to a local maxima. Panel (b) of figure \ref{fig:dendro_algo} also shows that the reference value has now encountered a local maxima to the right of the first detected structure and therefore, the algorithm assigns this peak to be a new structure. The reference value moves further down and both the structures grow by joining their corresponding neighbouring pixels until the reference value encounters a local minima between the two structures as seen in panel (c). At this stage, the two local maxima are merged together into a `branch'. The two local maxima are now treated as `leaves' of this branch. As the reference value moves further down, it encounters two new local maxima which are identified as new structures. On encountering a local minima between any two adjacent structures, the pre-existing structures are again merged similarly as before leading to the creation of a structure tree as seen in panel (d) of figure \ref{fig:dendro_algo}. The reference value stops at the noise level below which any existing structures are neglected. Finally, the structures in the structure tree (green lines shown in panel (d)) are numbered such that each structure can be identified by an unique structure ID. This simplified one dimensional scenario can be extended analogously to a three dimensional data set which will correspondingly detect three dimensional structures (pixels in x, y and z directions).

It has been known from observations that FTEs are generally enclosed within a region of higher thermal-pressure than the ambient, albeit with a decreasing pressure profile towards the core \citep{Hasegawa_2006, AkhavanTafti_2019}. Several global-MHD as well as embedded-PIC (EPIC) simulations of the magnetosphere, however, have repeatedly shown in slices through the FTEs that such structures generally correspond to high pressure globules (localised regions of higher thermal pressure) \citep{Chen_2017, FarinasPerez_2018, Sun_2019} in the simulations. We therefore leverage this feature as a detection strategy for the volumetric identification of FTEs in our simulation. We provide a 3D data cube of the thermal pressure in the global-MHD domain as an input to the hierarchical structure finding algorithm.
\begin{figure}
    \centering
    \includegraphics[width= \linewidth]{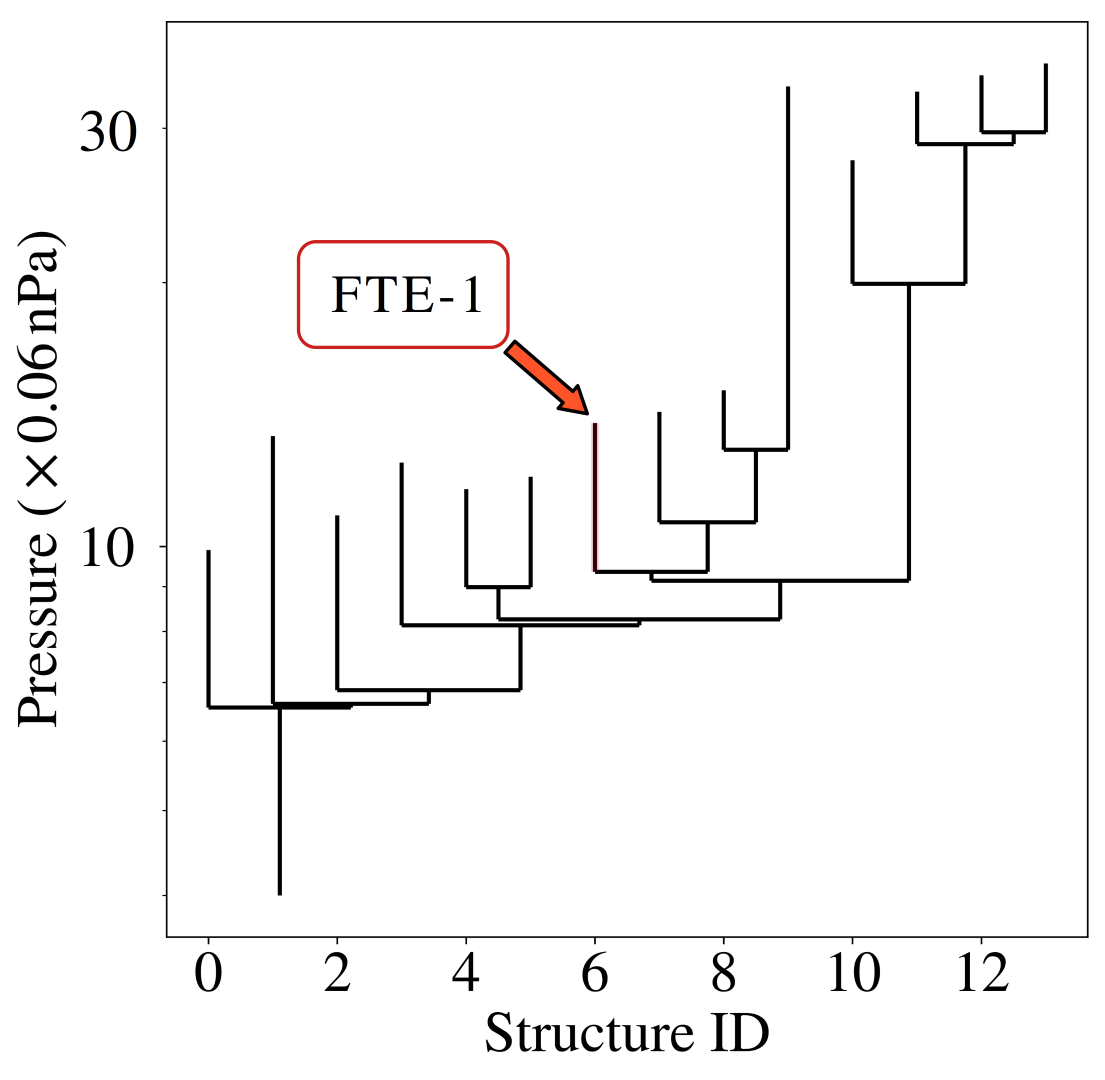}
    \caption{Plot shows a typical dendrogram obtained from the agglomerative hierarchical structure detection algorithm. The abscissa shows an unique structure ID for each detected structure and the ordinate corresponds to the magnitude of the volumetric pixels, i.e pressure in this case. The structure corresponding to FTE-1 at $\rm t_{det}\sim1.3\,min$ is also highlighted.}
    \label{fig:dendrogram}
\end{figure}
The noise threshold for the detection is dynamically set ascertaining that it is at-least 0.1 nPa higher than the ambient background thermal pressure. The exact value of the threshold is inconsequential in this regard as long as it is significantly lower than the thermal pressure of the FTE bulge. A threshold of $\sim$0.3 nPa generally appears to be sufficient for our detection. To volumetrically detect a particular FTE, say FTE-1 from the output data of the simulation, we first identify the FTE in the plots for $B_N$ at the magnetopause surface for each output data file. Each output data file is obtained at a particular time snapshot of the simulation and any two consecutive output files are temporally spaced by $\rm \sim 15$ seconds. Each of the output data files where FTE-1 can be identified by the bipolar $B_N$ signature is then isolated (9 consecutive data files) and a 3D data cube of the thermal pressure from each file is passed on to the agglomerative hierarchical clustering algorithm.

Figure \ref{fig:dendrogram} depicts a typical output from the agglomerative hierarchical clustering algorithm. Such an output is obtained for a particular temporal snapshot (one of the data files containing the FTE). The abscissa corresponds to the structure ID which is an unique identifier for each detected structure whereas the ordinate depicts the magnitude of the volumetric pixels, i.e, the value of the corresponding thermal pressure in this case. For each data file, the algorithm classifies the entire magnetosphere as a ``trunk" in the hierarchy, on which, the ``branches" correspond to a broad region surrounding the FTEs. The ``leaves" of the structure tree are the prime focus as they correspond to the volume enclosed by the FTEs in the data cube. The output dendrogram is pruned to get rid of other high-pressure regions in the domain that are not of interest to us such as, the polar cusps, as well as a section of the magnetotail region within the domain. The detected FTE structures generally have an azimuthal extent with tapered ends and an irregular cross section along this length thereby enclosing a closed volume. Each structure ID in the structure hierarchy is linked to the grid cells that constitute the structure.

Special care was taken to ensure that the selected structures indeed correspond to FTEs by manually examining the volumes of each of the detected FTEs throughout their time evolution. For each data file where an FTE is identified, a validation exercise similar to that described in section \ref{sec:FTE_det_validation} has been carried out. Section \ref{sec:FTE_det_validation} is a segment dedicated to elaborate the validation procedure employed for ascertaining that the output structures from the clustering algorithm indeed accurately correspond to the FTEs under consideration. Once the structure is properly isolated, the FTE volume can be calculated from the number of grid cells constituting that structure and the coordinate locations of the constituent grid cells can be used to retrieve any associated quantities at that location.

As the spatial resolution employed in this study is very high, the FTEs tend to be well resolved and are also often seen to be composed of multiple sub-structures. To avoid the detection of these FTE sub-structures as ``leaves" of the dendrogram, the input pressure data cube was first passed through a multidimensional Gaussian-filter. We have ascertained during the manual inspection process that the Gaussian-filter does not alter the integrity of the detection.  Finally, we halt the detection for each FTE when they are far enough from the subsolar point and are either (a) within $\rm \sim 0.5 R_E$ of the polar cusps or (b) out of the finest refinement level or (c) within $\rm \sim 0.5 R_E$ of the Kelvin-Helmholtz vortices forming at the flanks.

\end{document}